\documentclass[twocolumn,amsmath,amssymb,prd]{revtex4}

\def\al{\alpha}
\def\be{\beta}
\def\ga{\gamma}
\def\de{\delta}

\def\et{\eta}
\def\th{\theta}

\def\la{\lambda}

\def\si{\sigma}

\def\ch{\chi}
\def\ps{\psi}
\def\om{\omega}

\def\De{\Delta}

\def\La{\Lambda}

\def\Om{\Omega}

\def\cH{{\cal H}}
\def\cl{{\cal L}}

\def\fr#1#2{{{#1}\over{#2}}}
\def\frac#1#2{{\textstyle{{#1}\over{#2}}}}
\def\half{{\textstyle{1\over 2}}}
\def\ol{\overline}
\def\prt{\partial}

\def\lsim{\mathrel{\rlap{\lower4pt\hbox{\hskip1pt$\sim$}}
    \raise1pt\hbox{$<$}}}
\def\gsim{\mathrel{\rlap{\lower4pt\hbox{\hskip1pt$\sim$}}
    \raise1pt\hbox{$>$}}}

\def\etal{{\it et al.}}
\def\nn{\nonumber}

\def\vev#1{\langle {#1}\rangle}

\newcommand{\beq}{\begin{equation}}
\newcommand{\eeq}{\end{equation}}
\newcommand{\bea}{\begin{eqnarray}}
\newcommand{\eea}{\end{eqnarray}}
\newcommand{\rf}[1]{(\ref{#1})}
\newcommand{\eq}[1]{Eq.~(\ref{#1})}

\newcommand{\Tab}[1]{Table~\ref{#1}}
\newcommand{\Sec}[1]{Section~\ref{#1}}

\def\psb{\ol\ps{}}
\def\mbf#1{\boldsymbol #1}

\def\pvec{\mbf p}
\def\gavec{\mbf\ga}

\def\Q{\mathcal Q}

\def\Qhat{\widehat\Q}

\def\X{X}
\def\Y{Y}

\def\Xhat{\widehat\X}
\def\Yhat{\widehat\Y}

\def\cmtemplate#1#2#3#4{{#1}^{#3}_{#4}}

\def\ctemplate#1#2#3#4{{#1}^{(#2)#3}_{#4}}

\def\bcf#1#2{\ctemplate{b}{#1}{#2}{F}}

\def\cmtemplate#1#2#3#4{{#1}^{#3}_{#4}}

\def\bcmw#1#2#3{\cmtemplate{b}{#1}{#2}{{#3}}}

\def\dcmw#1#2#3{\cmtemplate{d}{#1}{#2}{{#3}}}

\def\gcmw#1#2#3{\cmtemplate{g}{#1}{#2}{{#3}}}
\def\Hcmw#1#2#3{\cmtemplate{H}{#1}{#2}{{#3}}}

\def\ctemplate#1#2#3#4{{#1}^{(#2)#3}_{#4}}

\def\bcw#1#2#3{\ctemplate{b}{#1}{#2}{{#3}}}

\def\dcw#1#2#3{\ctemplate{d}{#1}{#2}{{#3}}}

\def\gcw#1#2#3{\ctemplate{g}{#1}{#2}{{#3}}}
\def\Hcw#1#2#3{\ctemplate{H}{#1}{#2}{{#3}}}

\def\bcfw#1#2#3{\ctemplate{b}{#1}{#2}{F{,#3}}}

\def\dcfw#1#2#3{\ctemplate{d}{#1}{#2}{F{,#3}}}

\def\gcfw#1#2#3{\ctemplate{g}{#1}{#2}{F{,#3}}}
\def\Hcfw#1#2#3{\ctemplate{H}{#1}{#2}{F{,#3}}}

\def\ab{{\al\be}}

\def\mab{{\mu\al\be}}

\def\m{m_\ps}
\def\mw{m_w}

\def\btw#1#2{{\widetilde b}_{#1}^{#2}}
\def\bftw#1#2{{\widetilde b}_{F,#1}^{#2}}

\def\btws#1#2{{\widetilde b}_{#1}^{*#2}}
\def\bftws#1#2{{\widetilde b}_{F,#1}^{*#2}}

\def\bptw#1#2{{\widetilde b}_{#1}^{\prime #2}}

\def\ctw#1#2{{\widetilde c}_{#1}^{#2}}
\def\bptws#1#2{{\widetilde b}_{#1}^{\prime *#2}}

\def\ctws#1#2{{\widetilde c}_{#1}^{*#2}}

\def\bevec{\mbf\be}
\def\bE{\be_\oplus}
\def\beL{\be_E}

\def\vos{\mathrel{\rlap{\lower0pt\hbox{\hskip0.5pt{$\scriptstyle s$}}}
    \raise2pt\hbox{$\scriptstyle \neg$}}}

\begin{document}

\title{
Lorentz- and CPT-violating effects in Penning traps at linear boost
} 

\author{ 
Ariam J. Acevedo-L\'opez$^1$, 
Yunhua Ding$^2$, 
Kaito Iwasaki$^2$\footnote{
Present address: 
Department of Mathematics, University of Michigan, Ann Arbor, MI 48109, USA
}, 
and 
Arnaldo J.\ Vargas$^1$}

\affiliation{
$^1$Laboratory of Theoretical Physics, Department of Physics, University of Puerto Rico, R\'io Piedras, Puerto Rico 00936, USA\\
$^2$Department of Physics and Astronomy, Ohio Wesleyan University, Delaware, OH 43015, USA
}

\date{\today}

\begin{abstract}
We present in this work 
an analysis of Lorentz- and CPT-violating signals at linear boost order
in Penning-trap experiments.
The theory of quantum electrodynamics with Penning traps is revisited 
and the dominant shifts in the cyclotron and anomaly frequencies 
of confined particles and antiparticles are reproduced.
To study time variations of the experimental signals at linear boost order,
we provide a general discussion on transformations of coefficients for Lorentz violation 
between different frames,
and derive the expressions of the cyclotron and anomaly frequency shifts 
in the Sun-centered frame. 
Relating these frequency shifts to
the charge-to-mass ratios,
the $g$ factors,
and their comparisons between particles and antiparticles,
we extract numerous new or improved bounds on coefficients for Lorentz violation
from existing Penning-trap measurements.
 
\end{abstract}

\maketitle

\section{Introduction}
\label{introduction}

Invariance under Lorentz transformations stands as
a foundational symmetry of both 
general relativity and the Standard Model of particle physics.
However, 
tiny deviation of Lorentz symmetry could naturally emerge 
in a more fundamental theory that unifies gravity with quantum physics,
such as string 
theory~\cite{ksp}.
Since CPT violation in effective field theory 
is accompanied by Lorentz 
violation~\cite{ck,owg},
it follows that testing Lorentz symmetry includes CPT tests as well. 
Motivated by this,
numerous high-precision experiments spanning 
over various subfields of physics
have been performed to search for 
possible Lorentz- and CPT-violating 
signals~\cite{tables}.
Among these experiments, 
the Penning trap is of particular interest,
as it provides impressive sensitivities 
to measurements of the fundamental properties 
of particles and 
antiparticles~\cite{fa23, vd87, ga99, ul15, 17sc, 17sm, bo22},
permitting searches for any tiny deviation 
from Lorentz and CPT symmetry. 

A well-known signal for Lorentz and CPT violation is an observable 
that depends on the orientation and velocity of the experimental system 
relative to a fixed inertial reference frame. 
The standard reference frame used in the literature, 
known as the canonical Sun-centered frame \cite{sunframe}, 
has the property that the speed of the experimental system, 
typically located on the surface of the Earth, 
is small compared to the speed of light. 
This property is frequently exploited by expanding the observable 
as a power series of the small boost parameter 
obtained from the system's velocity 
and truncating the expansion to some order in the boost parameter 
to simplify the analysis. 
The most popular choice is to study the dominant effects by
truncating the expansion  at the zeroth boost order. 
This approach produces signals known as sidereal variations
and investigations of the sidereal variations have resulted in many 
high-precision tests of Lorentz and CPT symmetry.  
For example, 
Refs.~\cite{mi99,dk16, d19, 19sm, dr20, d21}
have studied the sidereal variations of 
Lorentz- and CPT-violating signals arising from 
Penning-trap experiments measuring the charge-to-mass ratios, 
the $g$ factors,
and their comparisons between particles and antiparticles.

Besides the success of this approximation, 
the limitation on effects at only the zeroth order in boost
excludes the possibility of identifying new types of
Lorentz- and CPT-violating signals 
linked to a change in the linear motion of the system. 
A prime example of a signal that becomes discernible at least at first boost order 
but not at the zeroth order is the annual variation of the observable, 
arising from the Earth's orbit around the Sun in the presence of Lorentz 
violation~\cite{sunframe,bklr03}. 
Publications considering contributions to the boost-dependent variation of an observable up to 
first~\cite{annual1,annual2, gkv14, kv18, kv15, torsion} 
or second boost 
order~\cite{bsquare,kmm16}
exist in the literature.
In particular, 
Ref.~\cite{kmm16} reported an apparent variation 
of the observable at the second harmonic of the annual frequency 
which is one of the signals for Lorentz violation predicted at the second boost order. 
However, 
no study of the effects at linear boost order in Penning-trap experiments 
has been performed to date.
Extending the Penning-trap analysis to include the effects at linear order in boost
is of significance as it can reveal additional types of measurable 
observables that cannot be studied by a sidereal variation analysis. 
 
In the absence of
compelling experimental evidence for Lorentz violation so far,
instead of constructing a specific model for Lorentz violation,
we take a more realistic approach by adopting
a general theoretical framework for Lorentz violation
to conduct a comprehensive study of possible effects. 
This framework is known as
the Standard-Model Extension 
(SME)~\cite{ck,akgrav},
which is developed in the context of effective field theory 
by adding all possible Lorentz-violating terms into 
the action of general relativity and the Standard Model.
Each of these terms is constructed from a coordinate-independent 
contraction between a Lorentz-violating operator and a corresponding controlling coefficient,
typically referred to the SME coefficient or coefficient for Lorentz violation.
The subset of the SME that restricts to power-counting renormalizable operators of 
mass dimensions $d\leq 4$ is called the minimal SME,
while the nonminimal SME contains operators of mass dimensions $d>4$
and is assumed to produce suppressed effects to conventional physics.

In this work,
within the SME framework,
we extend the previous studies of Lorentz- and CPT-violating effects 
in Penning-trap experiments by considering additional contributions at linear boost order.
Both the minimal and nonminimal SME produces various measurable 
Lorentz- and CPT-violating effects in Penning-trap experiments. 
These effects include shifts in the cyclotron and anomaly frequencies 
that can depend on time and also differ between particles and antiparticles.
To keep a reasonable scope of this work,
we restrict our attention to the effects due to the minimal SME.
The treatment of nonminimal SME effects at linear boost order  
would be an excellent subject of future work. 
To investigate the boost effects in Penning traps, 
results on the cyclotron and anomaly frequency shifts due to Lorentz and CPT violation 
obtained in 
Refs.~\cite{dk16} and \cite{dr20}  
provides a solid foundation. 
Applying the general Lorentz transformation at linear boost order,
we express the frequency shifts in terms of the SME coefficients in the Sun-centered frame
and study their time-dependence structure. 
Relating these expressions to the charge-to-mass ratios, the $g$ factors,
and their comparisons between particles and antiparticles in Penning-trap experiments, 
we extract constraints of the SME coefficients from available experimental results. 
The results derived from this work are complementary to the
existing ones from sidereal variation studies of Penning-trap 
effects~\cite{dk16, dr20, d21},
the investigations of the anomalous magnetic moment of muons in a storage 
ring~\cite{muon,gkv14},
the spectroscopic studies of hydrogen, antihydrogen, and other related 
systems~\cite{kv15},
and experiments involving clock 
comparisons~\cite{kv18}.

This work is organized as follows.
In 
\Sec{theory},
we revisit the theory of quantum electrodynamics 
with Lorentz- and CPT-violating operators of mass dimensions up to six
and its application to confined particles and antiparticles in Penning-trap experiments. 
The perturbative Hamiltonian at leading order in Lorentz and CPT violation
is presented in 
\Sec{the SME Lagrangian and Hamiltonian}.
The result is then applied in 
\Sec{perturbative energy shifts}
to discuss the dominant Lorentz- and CPT-violating  energy shifts 
of a confined particle or antiparticle in a Penning trap.
The shifts in the 
cyclotron and anomaly frequencies are reproduced in
\Sec{cyclotron and anomaly frequencies}.
We next address in 
\Sec{transformations} 
the general transformation of SME coefficients 
from the apparatus frame to the Sun-centered frame. 
Restricting our attention to the minimal SME,
we derive the expressions of the cyclotron and anomaly frequency shifts 
in terms of the Sun-centered frame SME coefficients at linear boost order.
We next turn in 
\Sec{experiments}
to applications to Penning-trap experiments involving confined protons and antiprotons.
We begin in
\Sec{the charge-to-mass ratios}
with a general discussion of the relationship of
the charge-to-mass ratio comparisons 
with the difference of cyclotron frequency shifts between protons and antiprotons.
The result is then applied to the ATRAP and BASE Penning-trap 
experiments to obtain limits on the cyclotron frequency shifts. 
Next, 
we consider in
\Sec{the g factors}
the applications to Penning-trap experiments that measure 
and compare the $g$ factors between protons and antiprotons. 
We first relate the $g$ factor comparisons between protons and antiprotons
to their anomaly frequency shifts,
and then apply it to the BASE experiments to derive relevant limits 
on the anomaly frequency shifts. 
To illustrate the process of extracting limits on the SME coefficients,
we provide in 
\Sec{results}
an explicit example using the BASE experiment comparing 
the charge-to-mass ratios between protons and antiprotons.
Finally, 
using published results from Penning-trap measurements, 
we obtain first-time constraints on 18 SME coefficients
and improve limits on 2 additional SME coefficients as well.
Some comments on the prospects of improving the current SME limits
or imposing more first-time SME limits are offered in 
\Sec{prospects}.
A summary of this work is provided in 
\Sec{summary}.
For completeness,
Appendix~\ref{appA}
presents the contributions to the cyclotron and anomaly frequency shifts
including effects at the zeroth order in the boost. 
Throughout the paper,
we follow the same notation used in 
Refs.~\cite{dk16} and \cite{dr20},
unless otherwise specified. 
Natural units with $\hbar =c = 1$
and mass units in GeV
are adopted throughout the paper.

\section{Theory}
\label{theory}

In this section,
we revisit the theory of Lorentz-violating spinor
electrodynamics with operators of mass dimensions up to six, 
which was developed in 
Ref.~\cite{dk16}.
By applying the theory to the Penning trap, 
we reproduce the leading-order contributions to
the cyclotron and anomaly frequencies
of confined particles and antiparticles
due to Lorentz and CPT violation.

\subsection{The SME Lagrangian and Hamiltonian}
\label{the SME Lagrangian and Hamiltonian}

The SME Lagrangian can be written as the conventional Standard Model Lagrangian 
plus all possible terms that break Lorentz symmetry.
For a single Dirac fermion field $\ps$ with charge $q$ and mass $\m$,
the Lagrangian density $\cl_\ps$
can be obtained by adding a general Lorentz-violating operator $\Qhat$  
to the conventional Lagrangian density,
\bea
\cl_\ps =
\half \psb (\ga^\mu i D_\mu - \m + \Qhat) \ps + {\rm H.c.} ,
\label{fermlag}
\eea
where
$D_\mu = (\prt_\mu + i q A_\mu)$ represents the covariant derivative 
involving the electromagnetic four-potential $A_\mu$ by the minimal coupling, 
and H.c. stands for Hermitian conjugate. 
The general Lorentz-violating operator $\Qhat$ in the Lagrangian density \rf{fermlag} 
contains terms formed by the contraction of coefficients for Lorentz violation, 
the covariant derivative $iD_\mu$, 
the antisymmetric electromagnetic field tensor 
$F_\ab \equiv \prt_\al A_\be - \prt_\be A_\al$, 
and one of the 16 Dirac basis matrices.   
For example, 
one of the dimension-five operators involving the $F$-type coefficients for Lorentz violation 
takes the form $\bcf 5 \mab F_\ab \ga_5 \ga_\mu $.
A comprehensive list of the relevant coefficients for Lorentz violation and their properties, 
up to mass dimensions $d\leq 6$, 
can be found in Table~I of 
Ref.~\cite{dk16}.
It is worth noting that the hermiticity of the Lagrangian density \rf{fermlag} 
requires that the operator $\Qhat$ satisfies the condition $\Qhat = \ga_0 \Qhat^\dag\ga_0$.
In the case of free fermions with $A_\al = 0$, 
the explicit expression of the Lagrangian density \rf{fermlag} at arbitrary mass dimension 
has been studied in 
Ref.~\cite{km13}. 
For the interaction case with $A_\al \neq 0$, 
Ref.~\cite{dk16}
developed a theory for operators with mass dimensions up to six.   
An extension of this theory to include operators of arbitrary mass dimension 
was recently presented in 
Ref.~\cite{kl19}.
Similar analyses have also been performed for other SME sectors,
including
photon~\cite{km09},
neutrino~\cite{km12},
and 
gravity~\cite{nonmingrav}.

The presence of the general operator $\Qhat$ 
in the Lagrange density \rf{fermlag}
modifies the conventional Dirac equation for a fermion in electromagnetic fields
to
\bea
\label{moddirac}
(p \cdot \ga - \m + \Qhat ) \ps = 0 ,
\eea
where $p_\al  = iD_\al$.
As no Lorentz-violating signals have been observed thus far, 
any such signal is expected to be extremely small compared 
to the energy scale of the system of interest. 
Consequently, 
we can treat the corrections due to Lorentz and CPT violation 
to the conventional Hamiltonian as perturbative
and apply perturbation theory to calculate the dominant shifts in the energy levels 
of the confined particles and antiparticles.
Based on the modified Dirac equation \rf{moddirac}, 
the exact Hamiltonian $\cH$ is defined as
\bea
\cH \ps \equiv p^0 \ps = \ga_0 (\pvec \cdot \gavec + \m - \Qhat)\ps=(\cH_0 + \de \cH)\ps,
\eea
where $p^0$ represents the exact energy of the system, 
encompassing all contributions from Lorentz and CPT violation,
$\pvec$ and $\gavec$ are the canonical momentum and gamma matrix vectors, respectively,  
$\cH_0$ denotes the conventional Hamiltonian for a fermion in an electromagnetic field, 
and
$\de \cH=-\ga_0 \Qhat$ 
represents the exact perturbative Hamiltonian.

To derive the perturbative Hamiltonian $\de \cH$,
we note that the operator $\Qhat$ generally contains terms that involve powers of $p^0$, 
corresponding to the exact Hamiltonian $\cH$ itself. 
In certain simple cases, 
it is possible to perform an appropriate field redefinition to eliminate 
the additional time derivatives and then adopt the standard procedure involving time translation 
on wave functions to obtain the exact perturbative Hamiltonian 
$\cH$~\cite{bkr98}.
However, 
in more general situations where powers of time derivatives exist, 
directly constructing $\de \cH$ becomes challenging. 
Nevertheless, 
we notice that any contributions to $\de \cH$ due to the exact Hamiltonian $\cH$ 
are at second order or higher in the coefficients for Lorentz violation.
Therefore, 
to obtain the leading-order results, 
one can apply the following substitution, 
as proposed in Refs. \cite{km12} and \cite{dk16},
\bea
\de\cH \simeq -\ga_0 \Qhat |_{p^0\to E_0},
\label{perturbH}
\eea
where $E_0$ is the unperturbed eigenvalue, 
which can be obtained by solving the conventional Dirac equation 
for a fermion in an electromagnetic field.

\subsection{Perturbative energy shifts}
\label{perturbative energy shifts}

Given the perturbative Hamiltonian $\de \cH$ as defined by expression  \rf{perturbH}, 
the Lorentz- and CPT-violating perturbative energy levels $\de E_{n, \pm}$ of a confined particle 
can be calculated using perturbation theory,
\bea
\label{deE}
\de E_{n, \pm}= \vev{\ch_{n, \pm}|\de\cH|\ch_{n, \pm}},
\eea
where $\chi_{n, \pm}$ denote the unperturbed stationary eigenstates, 
$n$ specifies the energy level number,
and $\pm$ represent the spin states of a positive-energy fermion. 

Before applying expression \rf{deE} to a confined particle in a Penning trap,
we note that a Penning trap can be idealized as 
a uniform magnetic field responsible for confining the radial motion of the particle
plus a quadrupole electric field providing confinement along the axial direction.
The primary contributions to the nonperturbative energy levels
are from the interactions of the confined particle with the magnetic field,
since the effects arising from the quadrupole electric field are suppressed
by a factor of $E/B\simeq 10^{-5}$ in natural units for a typical field configuration of
$E \simeq 20$~kV/m and $B \simeq 5$~T in a trap. 
Consequently, 
to obtain the dominant results in the energy shifts, 
we can simplify the trap configuration even further 
by conceptualizing it as consisting solely of a uniform magnetic field
in which a quantum fermion moves. 

The corresponding shifts in the energy levels $\de E^c_{n,\pm}$
due to Lorentz and CPT violation for antifermions
can be obtained by applying the perturbation theory in a similar way,
\bea
\label{deEs}
\de E^c_{n,\pm}= \vev{\ch^c_{n,\pm}|\de\cH^c |\ch^c_{n,\pm}} ,
\eea
where $\ch^c_{n,\pm}$ represent the eigenstates of positive-energy antifermions
obtained from the solutions of negative-energy fermions $\ch_{n,\pm}$ by charge conjugation,
$\de\cH^c$ is the perturbative Hamiltonian for the antifermion
derived from $\de\cH$ by charge conjugation in a similar way,
$n$ specifies the energy level number,
and $\pm$ denote the spin states for the antifermion as before.

\subsection{Cyclotron and anomaly frequencies}
\label{cyclotron and anomaly frequencies}

The primary observables of interest in a Penning-trap experiment are frequencies. 
Two key frequencies are the cyclotron frequency $\nu_c \equiv \om_c/2\pi$ and
the Larmor spin-precession frequency $\nu_L \equiv \om_L/2\pi$,
with their difference denoted by
the anomaly frequency $\nu_L-\nu_c = \nu_a \equiv  \om_a/2\pi$.
In the Lorentz-invariant scenario,
the charge-to-mass ratio and the $g$ factor
of a confined particle
moving in a Penning trap with a magnetic field strength $B$
are related to the cyclotron and anomaly frequencies by
\bea
\label{qm}
\dfrac{|q|}{m}   = \dfrac{\om_c}{B} ,
\eea  
 and
\bea
\label{g}
\fr {g}{2} = \fr {\om_L}{\om_c}
=
1+ \fr {\om_a}{\om_c},
\label{ratio}
\eea
respectively. 

A frequency can be viewed as the difference between different energy levels. 
For a confined fermion of flavor $w$ and charge sign~$\si$ in a Penning trap,
the cyclotron and anomaly frequencies can be defined as the
energy difference between the following energy levels 
\cite{bkr98},
\bea
\label{fermi}
\om_c^{w} 
=
 E_{1,\si}^{w}- E_{0,\si}^{w} ,
\qquad
 \om_a^{w} 
=
 E_{0,-\si}^{w}- E_{1,\si}^{w}.
\eea
The corresponding definitions for the cyclotron and anomaly frequencies 
of an antifermion of flavor $\ol{w}$ are given by 
\cite{bkr98}
\bea
\label{antifermi}
\om_c^{\ol{w}}
=
 E_{1,\si}^{\ol{w}}- E_{0,\si}^{\ol{w}} ,
\qquad
 \om_a^{\ol{w}} 
=
 E_{0,-\si}^{\ol{w}}- E_{1,\si}^{\ol{w}},
\eea
with the understanding that the charge signs $\si$ in definitions~\rf{antifermi}
are reversed compared to these in definitions~\rf{fermi}.

In the presence of Lorentz and CPT violation,
both the cyclotron and anomaly frequencies for fermions and antifermions 
can be shifted,
given by
\bea
\label{ca-lv-def}
\de \om_c^{w} 
=
\de E_{1,\si}^{w}- \de E_{0,\si}^{w} ,
\qquad
\de \om_a^{w} 
=
\de E_{0,-\si}^{w}- \de E_{1,\si}^{w},
\eea
for fermions, 
and 
\bea
\label{cas-lv-def}
\de \om_c^{\ol w} 
=
\de E_{1,\si}^{\ol w}-\de E_{0,\si}^{\ol w} 
\qquad
\de \om_a^{\ol w}
=
\de E_{0,-\si}^{\ol w}-\de E_{1,\si}^{\ol w},
\eea
for antifermions,
respectively.

Applying perturbation \rf{deE} and \rf{deEs} and following the definitions in expression \rf{ca-lv-def},
the cyclotron and anomaly frequency shifts of a fermion 
due to Lorentz violation are found to be
\cite{dr20}
\bea
\label{ca-lv}
\de \omega_c^{w}
&=&
\Big(
\dfrac{1}{m_w^2} \bptw w {3} 
- \dfrac{1}{m_w} (\ctw w {00} + \ctw w {11} + \ctw w {22})
\nn \\
&&
\quad - (\btw w {311} + \btw w {322}) 
\Big) eB,
\nn \\
\de \om_a^{w} 
&=&
 2 \btw w 3 - 2 \bftw w {33} B ,
\eea
where the tilde coefficients are given by
\bea
\label{tild}
\bptw w {3}
&=&
b_{w}^{3} + m_w (g_{w}^{120} - g_{w}^{012} + g_{w}^{021}) 
- m_w^2 b_{w}^{(5)300} 
\nn \\
&&
- 2 m_w^2 (H_{w}^{(5)1200} - H_{w}^{(5)0102} + H_{w}^{(5)0201})
\nn \\
&&
+ 2 m_{w}^3 d_{w}^{(6)3000}  
\nn \\
&&
+ 3 m_w^3 (g_{w}^{(6)12000}-g_{w}^{(6)01002}+g_{w}^{(6)02001}) ,
\nn \\
\ctw w {00}
&=&
c_{w}^{00} - m_w m_{w}^{(5)00} - 2 m_w a_{w}^{(5)000} 
\nn \\
&&
+ 3 m_w^2 c_{w}^{(6)0000} + 2 m_w^2 e_{w}^{(6)000} ,
\nn \\
\ctw w {jj} 
&=&
c_{w}^{jj} 
- 2 m_w a_{w}^{(5)j0j} 
+ 3 m_w^2  c_{w}^{(6)j00j} 
- m_w a_{w}^{(5)0jj} 
\nn\\
&&
- m_w m_{w}^{(5)jj} 
+ 3 m_w^2 c_{w}^{(6)00jj} 
+ 3 m_w^2 e_{w}^{(6)0jj}  ,
\nn\\
\btw w {3jj} 
& = &
b_{w}^{(5)3jj} + H_{w}^{(5)12jj} 
\nn \\
&&
- 3 m_{w} d_{w}^{(6)30jj} 
- 3 m_{w} g_{w}^{(6)120jj} ,
\nn \\
\btw w 3
&=&
\bcmw 3 3 w
+ \Hcmw 3 {12} w
- \mw \dcmw 4 {30} w
- \mw \gcmw 4 {120} w
+ \mw^2 \bcw 5 {300} w 
\nn \\
&&
+ \mw^2 \Hcw 5 {1200} w
- \mw^3 \dcw 6 {3000} w
- \mw^3 \gcw 6 {12000} w,
\nn\\
\bftw w {33}
&=&
\bcfw 5 {312} w
+ \Hcfw 5 {1212} w
- \mw \dcfw 6 {3012} w
- \mw \gcfw 6 {12012} w ,
\nn\\
\eea
with $j$ taking values of $1$ or $2$ (no summation assumed). 

In a similar way, 
taking the differences in expression \rf{cas-lv-def} gives the 
shifts in the cyclotron and anomaly frequencies of an antifermion 
due to Lorentz violation
\cite{dk16},
\bea
\label{cas-lv}
\de \omega_c^{\ol{w}}
&=&
\Big(
- \dfrac{1}{m_w^2} \bptws w {3} 
- \dfrac{1}{m_w} (\ctws w {00} + \ctws w {11} + \ctws w {22})
\nn \\
&&
\quad + (\btws w {311} + \btws w {322}) 
\Big) eB,
\nn \\
\de \om_a^{\ol w}
&=&
- 2 \btws w 3 + 2 \bftws w {33} B ,
\eea
where the starred tilde coefficients are given by
\bea
\label{tilds}
\bptws w {3}
&=&
b_{w}^{3} + m_w (g_{w}^{120} - g_{w}^{012} + g_{w}^{021}) 
- m_w^2 b_{w}^{(5)300} 
\nn \\
&&
+ 2 m_w^2 (H_{w}^{(5)1200} - H_{w}^{(5)0102} + H_{w}^{(5)0201})
\nn \\
&&
- 2 m_{w}^3 d_{w}^{(6)3000}  
\nn \\
&&
+ 3 m_w^3 (g_{w}^{(6)12000}-g_{w}^{(6)01002}+g_{w}^{(6)02001}) ,
\nn\\
\ctws w {00}
&=&
c_{w}^{00} - m_w m_{w}^{(5)00} + 2 m_w a_{w}^{(5)000} 
\nn \\
&&
+ 3 m_w^2 c_{w}^{(6)0000} - 2 m_w^2 e_{w}^{(6)000} ,
\nn \\
\ctws w {jj} 
&=&
c_{w}^{jj} 
+ 2 m_w a_{w}^{(5)j0j} 
+ 3 m_w^2  c_{w}^{(6)j00j} 
+ m_w a_{w}^{(5)0jj} 
\nn\\
&&
- m_w m_{w}^{(5)jj} 
+ 3 m_w^2 c_{w}^{(6)00jj} 
- 3 m_w^2 e_{w}^{(6)0jj}  ,
\nn\\
\btws w {3jj} 
& = &
b_{w}^{(5)3jj} - H_{w}^{(5)12jj} 
\nn \\
&&
+ 3 m_{w} d_{w}^{(6)30jj} 
- 3 m_{w} g_{w}^{(6)120jj} ,
\nn \\
\btws w 3
&=&
\bcmw 3 3 w
- \Hcmw 3 {12} w
+ \mw \dcmw 4 {30} w
- \mw \gcmw 4 {120} w
+ \mw^2 \bcw 5 {300} w
\nn \\
&&
- \mw^2 \Hcw 5 {1200} w
+ \mw^3 \dcw 6 {3000} w
- \mw^3 \gcw 6 {12000} w,
\nn\\
\bftws w {33}
&=&
\bcfw 5 {312} w
- \Hcfw 5 {1212} w
+ \mw \dcfw 6 {3012} w
- \mw \gcfw 6 {12012} w ,
\nn \\
\eea
with $j$ taking values of $1$ or $2$ (no summation assumed as before).

We note in passing that comparing the result~\rf{ca-lv} to \rf{cas-lv},
together with the relevant definitions 
\rf{tild} and \rf{tilds},
the shifts in the cyclotron and anomaly frequencies between a fermion and an antifermion 
differ only by the signs of all the basic coefficients for Lorentz violation that control CPT-odd effects,
as might be expected. 
We also remark in passing that 
the rotation properties of the coefficients for Lorentz violation in  
results \rf{ca-lv} and \rf{cas-lv}
are indicated by their indices.
To illustrate,
the pair of indices  ``12" on the right-hand sides of the definitions \rf{tild} and \rf{tilds}
are antisymmetric.  
In three dimensions, 
any antisymmetry pair of spatial indices rotate as a single spatial index. 
In particular, 
the antisymmetry indices ``12'' obey the same rotation rule as a single index ``3''.
This suggests that these particular coefficients for Lorentz violation 
undergo rotation transformations akin to a single index "3",
while coefficients with an index ``0" or a pair of indices ``00" are invariant under rotations. 
Also,
the cylindrical rotational symmetry inherent to the Penning trap is
correctly reflected in the fact that results
\rf{ca-lv} and \rf{cas-lv} only depend on 
index ``0", ``3", and ``11+22".
However, 
when considering boost transformations,   
each fundamental coefficient in 
definitions \rf{tild} and \rf{tilds}
has distinct transformation properties.

\section{Transformations}
\label{transformations}

The SME coefficients are assumed to be constant and uniform in any inertial reference frame.
The value of each coefficient is frame dependent as they transform as tensor components under observer transformations \cite{ck}, 
and in general, they are spacetime dependent in noninertial reference frames. 
For these reasons,  all the limits on SME coefficients should be 
reported in the same inertial reference frame to allow for any systematic comparison of the results obtained by different experiments. The canonical frame commonly adopted in the literature for this purpose is the Sun-centered celestial-equatorial frame \cite{sunframe}. By definition, the rest frame of the Sun is not an inertial reference frame, but it is more than close enough to one for our purpose. The origin of the Sun-centered frame is specified as the location of the Sun at the 2000 vernal equinox. The time coordinate $T$ is the cartesian coordinate time in the rest frame of the Sun. The spatial cartesian coordinates $X^J\equiv (X, Y, Z)$ are specified by aligning the $Z$ axis along the Earth's rotation axis and having the $X$ axis pointing from the Earth to the Sun at $T=0$. The $Y$ axis is obtained by completing a right-handed coordinate system.

After the preambles, we can move to the main part of this section that describes the Lorentz transformation used to express the frequency shifts in Eqs.~\eqref{ca-lv} and \eqref{cas-lv} in terms of the SME coefficients in the Sun-centered frame for an Earth-based experiment. It is convenient to separate the transformation into two stages. We start by transforming from the Sun-centered frame to the so-called standard laboratory frame with coordinates $x^{\mu}\equiv(t,x,y,z)$ \cite{sunframe}. 
The standard laboratory frame is instantaneously comoving with the laboratory, 
and its spatial axes are defined by having the $x$-axis pointing to the local south, the $y$-axis pointing to the local east, and the $z$-axis pointing to the local zenith. The final stage in the transformation is a rotation from the standard laboratory frame to the apparatus frame with cartesian coordinates $x^\mu \equiv (x^0, x^1, x^2, x^3)$. 

Eqs.~\eqref{ca-lv} and \eqref{cas-lv} are expressed in the apparatus frame that has the $x^3$ axis in the direction of the applied magnetic field \cite{dk16}. We can always define the orientation of the apparatus frame by specifying the Euler angles to rotate from the standard laboratory frame to the apparatus frame. Fortunately, all the systems considered in this work have their applied magnetic field parallel or perpendicular to a vector pointing toward the local zenith. We only need to introduce two convections for the apparatus frame depending on the relative orientation between the $z$ axis of the local standard laboratory frame and the magnetic field \cite{dk16}. For a vertical magnetic field, parallel to the $z$ axis, we define the apparatus frame as the standard laboratory frame. In other words, the rotation between the frames is the identity matrix with the coordinates related by $(x^0,x^1, x^2, x^3)=(t,x,y,z)$. For a horizontal magnetic field, perpendicular to the $z$ axis, we have the $x^2$ axis toward the local zenith, the $x^3$ axis in the direction of the applied field, and the $x^1$ axis obtained by the right-hand rule.

The observer Lorentz transformation $ \La^\mu_{\ \nu}({\mbf\th},\bevec)$ between the apparatus frame and the Sun-centered frame is the composition of a rotation $ \mathcal{R}^\mu_{\ \nu}({\mbf \th})$ with a boost $ \mathcal{B}^\mu_{\ \nu}(\bevec)$,
\beq
\La^\mu_{\ \nu}({\mbf\th},\bevec)=
\mathcal{R}^\mu_{\ \al}({\mbf \th})\mathcal{B}^\al_{\ \nu}(\bevec),
\label{lortr}
\eeq
where $\mbf \th$ is the rotation parameter, and $\bevec$ is the velocity of the apparatus frame in the Sun-centered frame. 
The speed $\be\simeq 10^{-4}$ between the frames is small compared to the speed of light. 
We can simplify the expression for $\La^\mu_{\ \nu}$ by expanding it as a power series of $\be$ and truncating it at some power of $\be$. The truncation of the power series to zeroth order in $\be$ reduces the transformation to a pure rotation, where the boost matrix in Eq. \eqref{lortr} is replaced with the identity matrix. Therefore, the Lorentz transformation $\La^{\mu}{}_{\nu}$ takes the form
\beq
\La^{0}{}_{T}=1,
\hskip 8pt
\La^{0}{}_{J}=
\La^{j}{}_{T}=0,
\hskip 8pt
\La^{j}{}_{J}=\mathcal{R}^{j}{}_{J},
\label{LT0}
\eeq
where lower-case and upper-case indices represent spatial cartesian coordinates in the apparatus frame and the Sun-centered frame,
respectively. The expressions for the Lorentz-violating frequency shifts 
in Eqs.~\eqref{ca-lv} and \rf{cas-lv} 
at the zeroth order in the boost were obtained in previous publications \cite{dr20,dk16, bkr98, bkr97}. 
Further below, we will reproduce some of the main results 
of these previous works to facilitate the discussion. Our goal in this work is to extend
 these previous works by expanding the Lorentz transformation in $\La^\mu_{\ \nu}$ to linear order in $\be$. At linear order in $\be$, we get that
\beq
\La^{0}{}_{T}=1,
\hskip 8pt
\La^{0}{}_{J}=-\bevec^J,
\hskip 8pt
\La^{j}{}_{T}=-\mathcal{R}^{j}{}_{J}\bevec^{J},
\hskip 8pt
\La^{j}{}_{J}=\mathcal{R}^{j}{}_{J}.
\label{LTlinear}
\eeq

It is convenient to introduce the local sidereal time $T_\oplus$ before discussing the main results of the previous works. The local sidereal time $T_\oplus$  is an offset from the time $T$ in the Sun-centered frame \cite{dk16},
\beq
T_\oplus\simeq T-\fr{(66.25^\circ- \la)}{360^\circ} 23.934~{\rm hr},
\label{T0}
\eeq
where $\la$ is the longitude of the laboratory in degrees. The crucial property of the sidereal time is that 
$\om_\oplus T_\oplus $
is a multiple of $2\pi$ every time that the $y$ axis in the standard laboratory frame lies along the $Y$ axis in the Sun-centered frame, where $\om_\oplus\simeq 2\pi/(23.934~{\rm hr})$ is the sidereal frequency of the Earth.

The rotation matrix $\mathcal{R}^{j}_{\ J}$ 
in Eqs. \rf{LT0} and \rf{LTlinear}
depends on the direction of the magnetic field. For a vertical magnetic field, it is given by \cite{sunframe,dk16}
\beq
\mathcal{R}^{j}_{\ J}=\left(
\begin{array}{ccc}
\cos\ch\cos\om_\oplus T_\oplus
&
\cos\ch\sin\om_\oplus T_\oplus
&
-\sin\ch
\\
-\sin\om_\oplus T_\oplus
&
\cos\om_\oplus T_\oplus
&
0
\\
\sin\ch\cos\om_\oplus T_\oplus
&
\sin\ch\sin\om_\oplus T_\oplus
&
\cos\ch
\end{array}
\right),
\label{rotmatV}
\eeq
where $\ch$ is the colatitude of the laboratory. For a horizontal magnetic field, the rotation matrix takes the form \cite{dk16}

\bea
\mathcal{R}^{j}_{\ J}&=&\left(
\begin{array}{ccc}
0
&
0
&
-1
\\
-\sin\th
&
\cos\th
&
0
\\
\cos\th
&
\sin\th
&
0
\end{array}
\right)\nn\\
&&\times
\left(
\begin{array}{ccc}
\cos\ch\cos\om_\oplus T_\oplus
&
\cos\ch\sin\om_\oplus T_\oplus
&
-\sin\ch
\\
-\sin\om_\oplus T_\oplus
&
\cos\om_\oplus T_\oplus
&
0
\\
\sin\ch\cos\om_\oplus T_\oplus
&
\sin\ch\sin\om_\oplus T_\oplus
&
\cos\ch
\end{array}\right),\nn\\
\label{rotmatH}
\eea
where $\th$ is the angle of the horizontal magnetic field from the local south assuming the convection that counter-clockwise angles are positive.  

The form of the frequency shifts in Eq.~\rf{ca-lv} 
in the Sun-centered frame, at the zeroth order in the boost, is obtained by expressing the apparatus-frame SME coefficients in terms of the Sun-centered-frame ones using Eq.~\eqref{LT0} together with Eq.~\eqref{rotmatV} or \eqref{rotmatH}. The effective coefficients in Eq.~\eqref{ca-lv}
 facilitate these transformations by grouping all the coefficients that transform similarly under rotations. 
 The effective coefficients $\btw{w}{j}$ and $\bptw{w}{j}$, 
 given in definition~\eqref{tild},
contain only the SME coefficients that rotate as vectors. All the coefficients that contribute to $\bftw{w}{jk}$ and $\ctw{w}{jk}$ rotate as rank-2 tensors, and the ones that contribute to $\btw{w}{jkl}$ as rank-3 tensors.

As an example, we will reproduce the results presented in Ref.~\cite{dk16} for the anomaly frequency assuming a vertical magnetic field.  The relevant effective coefficients transform as 
\bea
\btw w 3 &=&
\btw w Z \cos\ch
+ ( \btw w X \cos\om_\oplus T_\oplus
+ \btw w Y \sin\om_\oplus T_\oplus )\sin\ch , 
\nn\\
\eea
and
\bea
\bftw w {33}
&=&
\bftw w {ZZ}
+\half (\bftw w {XX} +\bftw w {YY} -2\bftw w {ZZ}) \sin^2\ch
\nn\\
&&
+( \bftw w {(XZ)} \cos\om_\oplus T_\oplus
+ \bftw w {(YZ)} \sin\om_\oplus T_\oplus )\sin 2\chi
\nn\\
&&
+[
\half (\bftw w {XX} - \bftw w {YY}) \cos 2\om_\oplus T_\oplus
\nn\\
&&
\hskip 45pt
+ \bftw w {(XY)} \sin 2\om_\oplus T_\oplus ] \sin^2\ch .
\eea
Applying these results in Eq.~\eqref{ca-lv} reveals that the Lorentz-violating anomaly frequency shifts of a particle at the zeroth order in $\be$ have the form
\bea
\de \om_{a,0}^w
&=&
A_0^{(a,0)}+A_c^{(a,0)} \cos\om_\oplus T_\oplus+A_s^{(a,0)} \sin\om_\oplus T_\oplus\nn\\
&&+A_{c2}^{(a,0)} \cos 2\om_\oplus T_\oplus+A_{s2}^{(a,0)} \sin 2\om_\oplus T_\oplus,
\label{dwa0m}
\eea
where the 0 subscript in $\de \om_{a,0}^w$ indicate the zeroth boost order.
The amplitudes  
$A^{(a,0)}_{\ast}$
are linear combinations of the SME coefficients
with subscripts ${\ast}$ ranging over values 
$0, c, s, c2, s2$ 
that specify the harmonics associated with the amplitudes.
According to this result, a signal for Lorentz violation is a sidereal variation of the anomaly frequency resulting from the rotation of the Earth relative to a fixed inertial reference frame. A sidereal variation of a resonance frequency is the most common signal for Lorentz violation studied in the literature \cite{tables}. The signals for Lorentz violation resulting from this analysis, at the zeroth order in the boost, were studied in detail in Refs. \cite{dk16,dr20}. A reproduction of the main expressions for the frequency shifts at the zeroth order in $\be$ obtained in these publications is listed in Appendix \ref{appA}. 
The analysis in Ref.~\cite{dk16} predicted a sidereal variation of the anomaly frequency with the first and second harmonic of the sidereal frequency, 
and the prediction of Ref.~\cite{dr20} is a variation of the cyclotron frequency up to the third harmonic of the sidereal frequency. The variation of the 
anomaly frequency is only with the first harmonic of the sidereal frequency and for the cyclotron frequency up to the second harmonic if we limited the scope of these works to the minimal SME coefficients by using the frequency shifts defined in 
\eq{ca-lv-min}.

After summarizing the previous works, we consider the advantages of including corrections
 at linear order in the boost. A drawback of limiting the frame transformation to zeroth order in $\be$ is that it disregards the contributions from some SME coefficients to the frequency shifts \cite{kv15}. The expansion of these previous works to linear order in the boost has the advantage of revealing a greater number of SME coefficients that can produce signals for Lorentz violation detectable in Penning-trap experiments. Another feature of expanding the analysis is to unveil new signals for Lorentz violation including an annual variation of the anomaly and cyclotron frequencies.
 
An implication from Eq.~\eqref{LTlinear} is that the time interval measured between events happening at the laboratory 
is the same in the apparatus frame as
in the Sun-centered frame at the first order in $\be$. From now on, we will express the time dependence of the Lorentz-violating frequency shifts using time intervals $\De T$ in the Sun-centered frame as they are identical, up to the first order in $\be$, to the time intervals measured in the laboratory frame.  The velocity $\bevec$ of the apparatus frame relative to the Sun-centered frame is approximately given by
\beq
\bevec\simeq\bevec_\oplus +\bevec_L,
\label{boostpar}
\eeq
where $\bevec_\oplus$ is the velocity of the Earth relative to the Sun
and $\bevec_\oplus$ is the velocity of the laboratory relative to Earth's center of mass,
both expressed in the Sun-centered frame. 
Taking the Earth's orbit as circular, we get that
\beq
\bevec_\oplus
=
\bE \sin{\Om_\oplus T} ~\widehat{X}
-\bE \cos{\Om_\oplus T}
( \cos\et~\widehat{Y} + \sin\et~\widehat{Z} ) ,
\label{vorb}
\eeq
where $\bE\simeq 10^{-4}$ is the Earth's orbital speed,
$\Om_\oplus\simeq 2\pi/(365.26 \text{ d})$
is the Earth's orbital angular frequency,
and $\et\simeq23.4^\circ$ is the angle
between the $XY$ plane and the Earth's orbital plane.
Treating the Earth as a sphere, we have
\beq
\bevec_L
=
r_\oplus \om_\oplus \sin{\ch}\left(-\sin{\om_\oplus T_\oplus}~\Xhat
+\cos{\om_\oplus T_\oplus}~\Yhat\right) ,
\label{vrot}
\eeq
where $\ch$ is again the colatitude of the laboratory,
$r_\oplus$ is the radius of the Earth, and $\om_\oplus$ is the sidereal frequency. 
The magnitude of $\be_L$ is around $10^{-6}$ and two orders of magnitude smaller than $\bE$. 
Note that the sidereal time $T_\oplus$
is used in Eq.~$\eqref{vrot}$
and the difference of  $T-T_\oplus$
given in \eq{T0}
 is a phase
that physically represents a convenient choice of a local time zero. 

A straightforward extension of the previous works would include contributions to the frequency shifts at linear order in $\be$ due to the SME coefficients with mass dimensions up to six. In this work, we opted for a more pragmatic approach by limiting the scope of our work to contributions at linear order in $\be$ due to the minimal Lorentz-violating operators. The challenge of including the nonminimal terms is that the expressions for the frequency shifts can become overwhelming as some of the coefficients that contribute transform as rank-5 tensors under observer Lorentz transformation even if they only transform as rank-3 tensors under rotations. Another justification to pursue this approach is that it results in new limits on previously unconstrained minimal SME coefficients as discussed in Sec. \ref{experiments}. 
The treatment of nonminimal SME coefficients at linear order in $\be$
would be a subject of future work. 
Keeping only the minimal SME coefficients in the cyclotron and anomaly frequency shifts~\rf{ca-lv} and \rf{cas-lv}, 
we have
\bea
\label{ca-lv-min}
\de \omega_c^{w}
&=&
\Big(
\dfrac{1}{m_w^2} \bptw w {3} 
- \dfrac{1}{m_w} (c_{w}^{00} + c_{w}^{11} + c_{w}^{22})
\Big) eB ,
\nn\\
\de \om_a^{w} 
&=&
 2 \btw w 3  ,
\eea
where the tilde coefficients are defined by
\bea
\label{tild-min}
\bptw w {3}
&=&
b_{w}^{3} + m_w (g_{w}^{120} - g_{w}^{012} + g_{w}^{021}) ,
\nn \\
\btw w 3
&=&
\bcmw 3 3 w
+ \Hcmw 3 {12} w
- \mw \dcmw 4 {30} w
- \mw \gcmw 4 {120} w ,
\eea
and
\bea
\label{cas-lv-min}
\de \omega_c^{\ol{w}}
&=&
\Big(
- \dfrac{1}{m_w^2} \bptws w {3} 
- \dfrac{1}{m_w} (c_{w}^{00} + c_{w}^{11} + c_{w}^{22}) 
\Big) eB,
\nn \\
\de \om_a^{\ol w}
&=&
- 2 \btws w 3  ,
\eea
where the starred tilde coefficients are given by
\bea
\label{tilds-min}
\bptws w 3
&=&
b_{w}^{3} + m_w (g_{w}^{120} - g_{w}^{012} + g_{w}^{021}) ,
\nn \\
\btws w 3
&=&
\bcmw 3 3 w
- \Hcmw 3 {12} w
+ \mw \dcmw 4 {30} w
- \mw \gcmw 4 {120} w.
\eea
Note $\bptws w 3$ and $\bptw w 3$ have the same expression in the limit of the minimal SME.

The notation in the cyclotron and anomaly frequency shifts 
\rf{ca-lv-min}
can be misleading if we move beyond the pure rotation approximation. For instance, the coefficient $\btw{w}{j}$ doesn't transform as a Lorentz vector under observer transformations. We can observe from Eq.~\eqref{tild-min} that $\btw{w}{j}$ is a linear combination of coefficients that transform differently from each other under observer Lorentz transformations. To illustrate the point, consider $b^3_w$ and $g_w^{120}$ that are two of the coefficients contained in $\btw{w}{3}$. The former coefficient transforms as a Lorentz vector while the latter transforms as a rank-3 Lorentz tensor. Hence, obtaining the expression for the frequency shifts in the Sun-centered frame requires abandoning the effective-coefficient notation and expressing the frequency shifts in terms of coefficients that transform as Lorentz tensors.
 
The SME coefficients in the apparatus frame can be expressed in terms of the SME coefficients in the Sun-centered frame at linear order in $\be$ by applying the Lorentz transformation \rf{lortr} together with the boost velocities \rf{boostpar}, \rf{vorb}, and \eqref{vrot}, and the rotation matrix \rf{rotmatV} or \rf{rotmatH}. For example, assuming a vertical magnetic field, the coefficient $b^3_w$ contained in $\btw{3}{w}$ transforms as
\bea
b^3_w
&=&
b^Z_w \cos\ch+\sin\ch(b^X_w\cos\om_\oplus T_\oplus+b^Y_w\sin\om_\oplus T_\oplus)
\nn\\
&&
+b^T_w\bE\cos\ch \sin\et \cos\Om_\oplus T
\nn\\
&&
-b^T_w\bE \sin\chi\cos\om_\oplus T_\oplus\sin\Om_\oplus T 
\nn\\
&&
+b^T_w\bE \sin\chi\cos\et  \sin \om_\oplus T_\oplus\cos\Om_\oplus T.
\label{btranf}
\eea
As previously stated, 
the coefficient $b^T_w$ appearing at linear order in the boost is independent of the coefficients 
$b^X_w$, $b^Y_w$, and $b^Z_w$ appearing at the zeroth order in $\be$. 

Keeping terms up to linear order in $\be$,
the Lorentz-violating cyclotron frequency shifts in
\eq{ca-lv-min} expressed in terms of the SME coefficients in the Sun-centered frame take the form 
\beq
\de \om_c^w\simeq \de \om_{c,0}^w + \de \om_{c,1}^w,
\eeq
where $\de \om_{c,0}^w$ are the boost-independent cyclotron frequency shifts 
and $\de \om_{c,1}^w$ denote the contributions at linear order in the boost.
The SME coefficients in $\de \om_{c,0}^w$ including nonminmal ones up to mass dimension six have been studied in detail
in Ref.~\cite{dr20}.
For completeness, 
we reproduce the zeroth order results $\de \om_{c,0}^w$ 
in \eq{dwc0} in 
Appendix \ref{appA}.
The ratio between $\de \om_{c,1}^w$ at linear order in $\be$ 
and the product of the unit electric charge $e$ and the magnetic field $B$ takes the form  
\bea
\fr{\de \om_{c,1}^w}{eB}
&=&
A_0^{(c,1)}+A_c^{(c,1)} \cos\om_\oplus T_\oplus+A_s^{(c,1)} \sin\om_\oplus T_\oplus
\nn\\
&&
+A_C^{(c,1)} \cos\Om_\oplus T+A_S^{(c,1)} \sin\Om_\oplus T
\nn\\
&&
+\cos \om_\oplus T_\oplus \left(A_{cC}^{(c,1)} \cos\Om_\oplus T+A_{cS}^{(c,1)} \sin\Om_\oplus T\right)
\nn\\
&&
+\sin \om_\oplus T_\oplus \left(A_{sC}^{(c,1)} \cos\Om_\oplus T+A_{sS}^{(c,1)} \sin\Om_\oplus T\right)
\nn\\
&&
+\cos 2\om_\oplus T_\oplus \left(A_{c2C}^{(c,1)} \cos \Om_\oplus T+A_{c2S}^{(c,1)} \sin \Om_\oplus T\right)
\nn\\
&&
+\sin 2\om_\oplus T_\oplus \left(A_{s2C}^{(c,1)} \cos \Om_\oplus T+A_{s2S}^{(c,1)} \sin \Om_\oplus T\right)
\nn\\
&&
+A_{c2}^{(c,1)} \cos 2\om_\oplus T_\oplus+A_{s2}^{(c,1)} \sin 2\om_\oplus T_\oplus ,
\label{dwc1}
\eea
where notation $A_{\ast}^{(c,1)}$ is used for the amplitude
of each harmonic, 
with superscripts $(c,1)$ representing the cyclotron frequency shifts at linear order in $\be$
and subscript~$\ast$ taking values ranging over $0, c, s, C, S, cC, cS, sC, sS, ...$
We list in Table \ref{cV} and Table \ref{cH}
the explicit expressions of the amplitudes $A_{\ast}^{(c,1)}$
for a vertical and horizontal magnetic field,
respectively. 
In each table, 
the first column specifies the amplitudes $A_{\ast}^{(c,1)}$. 
The second column lists the corresponding boost factors with $\bE\simeq 10^{-4}$ denoting 
the Earth's revolution velocity about the Sun 
and $\beL\equiv r_\oplus\om_\oplus\simeq 1.6\times 10^{-6}$ 
specifying the tangential velocity of a point on the Earth's equator due to the Earth's rotation, 
respectively.   
Finally, 
the third column gives the combinations of the Sun-centered frame SME coefficients. 
In the final column, notations $c_\vartheta =\cos\vartheta$ and $s_\vartheta=\sin\vartheta$ are used, 
where $\vartheta$ can represent the colatitude $\ch$, the angle $\et\simeq23.4^\circ$, 
or the angle $\th$ between the local south and the magnetic field. 
The amplitudes $A_{\ast}^{(c,1)}$  are obtained by multiplying the terms in the second and third columns.

\renewcommand{\arraystretch}{1.5}
\begin{table*}
\caption{
\label{cV}
SCF expressions of cyclotron frequency shifts for a vertical $\mbf B$.}
\setlength{\tabcolsep}{5pt}
\begin{tabular}{ccc}
\hline
\hline																							
Amplitude		          &		Boost 	         &		Coefficient 			\\	
	 		          &		factor			&		combination		 	\\	\hline
$	A_0^{(c,1)}	  $	&	$	\beL		$	&	$	 -\left(2 g_w^{TZT}+g_w^{XZX}+g_w^{YZY}\right) s^2_\chi/m_w		$	\\	
$	A_{c}^{(c,1)}     $	&	$	\beL		$	&	$	2\left(2c_w^{(TY)}+(g_w^{TXT}-g_w^{XYY}) c_\chi\right)s_\chi/m_w		$	\\	
$	A_{s}^{(c,1)}	  $	&	$	\beL		$	&	$	 2 \left((g_w^{TYT}+g_w^{XYX}) c_\ch-2c_w^{(TX)}\right) s_\ch/m_w		$	\\	
$	A_{C}^{(c,1)}	  $	&	$	\bE		$	&	$	\left[2 s_\et \left(\left(b_w^{T}+(g_w^{XYZ}+g_w^{XZY}-g_w^{YZX}) m_w\right) c_\chi+ c_w^{(TZ)} m_w (c_{2\chi}-3)\right)\right.		$	\\	
$		            $	&	$			$	&	$	  \left.+ c_\et m_w\left(4(g_w^{XYY}-g_w^{TXT}) c_\ch -c_w^{(TY)}(7+c_{2\chi})\right)\right]/2 m_w^2	$	\\	
$	A_{S}^{(c,1)}	  $	&	$	\bE		$	&	$	\left(c_w^{(TX)} (7+c_{2\chi})-4 (g_w^{TYT}+g_w^{XYX}) c_\chi\right)/2 m_w		$	\\	
$	A_{c2}^{(c,1)}   $	&	$	\beL		$	&	$	\left(g_w^{XZX}-g_w^{YZY}\right) s^2_\chi/m_w		$	\\	
$	A_{s2}	^{(c,1)}   $	&	$	\beL		$	&	$	\left(g_w^{XZY}+g_w^{YZX}\right) s^2_\chi/m_w		$	\\	
$	A_{cC}^{(c,1)}	  $	&	$	\bE		$	&	$	2\left(c_\et (g_w^{TZT}+g_w^{YZY})+ s_\et(g_w^{YZZ}-g_w^{TYT}+c_w^{(TX)} c_\chi)\right) s_\ch/m_w		$	\\	
$	A_{cS}^{(c,1)}	  $	&	$	\bE		$	&	$	-\left(b_w^{T}+(g_w^{XZY}-g_w^{XYZ}+g_w^{YZX}) m_w+2c_w^{(TZ)} m_w c_\ch\right) s_\ch/m_w^2		$	\\	
$	A_{sC}^{(c,1)}	  $	&	$	\bE		$	&	$	\left[c_\eta \left(b_w^{T}- m_w(g_w^{XYZ}+g_w^{XZY}+g_w^{YZX}-2c_w^{(TZ)}c_\chi)\right)+2 m_w s_\et  \left(g_w^{TXT}-g_w^{XZZ}+c_w^{(TY)} c_\chi\right)\right] s_\chi/m_w^2	$	\\	
$	A_{sS} ^{(c,1)}	  $	&	$	\bE		$	&	$	2 (g_w^{TZT}+g_w^{XZX}) s_\chi/m_w		$	\\	
$   A_{c2C}^{(c,1)}      $	&	$	\bE		$	&	$	 -c_w^{(TY)} c_\et s^2_\ch/m_w		$	\\	
$   A_{c2S}^{(c,1)}      $	&	$	\bE		$	&	$	 -c_w^{(TX)} s^2_\ch/m_w		$	\\	
$   A_{s2C}^{(c,1)}      $	&	$	\bE		$	&	$	c_w^{(TX)} c_\et s^2_\ch/m_w		$	\\	
$   A_{s2S}^{(c,1)}      $	&	$	\bE		$	&	$	 -c_w^{(TY)} s^2_\ch/m_w		$	\\			
\hline
\hline
\end{tabular}
\end{table*} 

\renewcommand{\arraystretch}{1.5}
\begin{table*}
\caption{
\label{cH}
SCF expressions of cyclotron frequency shifts for a horizontal $\mbf B$.}
\setlength{\tabcolsep}{1pt}
\begin{tabular}{ccc}
\hline
\hline																							
Amplitude		          &		Boost 	         &		Coefficient 	combination		\\	
	 		          & 		factor	                   &		combination		 	\\	\hline
$    A_0^{(c,1)}	  $	&	$	\beL	          $	&	$        s_\chi\left[ (g_w^{XYZ} m_w -b_w^{T})s_\th +\left(2c_w^{(TZ)} s_\th s_\ch-(2 g_w^{TZT}+g_w^{XZX}+g_w^{YZY}) c_\ch \right)m_w c_\th\right]/m_w^2  $	\\	
$   A_{c}^{(c,1)}         $	&	$	\beL		$	&	$	 s_\chi \left[ c_w^{(TY)} (3+c_{2\th})-c_w^{(TX)} c_\ch s_{2\th}-2 (g_w^{TXT}-g_w^{XYY}) c_\th s_\ch\right]/m_w		$	\\	
$    A_{s}^{(c,1)}	  $	&	$	\beL		$	&	$	- s_\chi \left[c_w^{(TX)} (3+c_{2\th})+c_w^{(TY)} c_\ch s_{2\th}+2 (g_w^{TYT}+g_w^{XYX}) c_\th s_\ch\right]/m_w	$	\\	
$    A_{C}^{(c,1)}	  $	&	$	\bE		$	&	$	\left[ m_w c_\et \left( 2 (g^{TXT}_w-g^{XYY}_w) c_\th s_\ch- c_w^{(TY)}(2+c_\th^2+c_\ch^2 s_\th^2+s_\ch^2)\right) -2m_w c_w^{(TZ)}s_\et(1+s_\th^2 s_\ch^2+c_\ch^2)\right.	$	\\	
$		            $      &	$			$        &	$	 \left.- \left( b^T_w + (g^{XYZ} _w + g^{XZY} _w - g^{YZX} _w) m_w\right) s_\et c_\th s_\ch\right]/ m_w^2		$	\\	
$    A_{S}^{(c,1)}	  $	&	$	\bE		$	&	$	\left[2 (g_w^{TYT}+g_w^{XYX}) c_\th s_\ch+c_w^{(TX)} (2+ c^2_\ch s^2_\th+s^2_\chi+c^2_\th)\right]/m_w		$	\\	
$    A_{c2}^{(c,1)}	  $	&	$	\beL		$	&	$	s_\ch\left[(g_w^{XZX}-g_w^{YZY}) c_\th c_\ch+(g_w^{XZY}+g_w^{YZX}) s_\th\right]/m_w			$	\\	
$    A_{s2}	^{(c,1)}   $	&	$	\beL		$	&	$	s_\ch \left[(g_w^{XZY}+g_w^{YZX}) c_\th c_\ch+(g_w^{YZY}-g_w^{XZX}) s_\th\right]/m_w		$	\\	
$    A_{cC}^{(c,1)}	  $	&	$	\bE		$	&	$	\left[  c_\et \left( \left( b^T_w - m_w(g^{XYZ}_w+g^{XZY}_w+g^{YZX})\right)s_\th+2 m_w  c_\th\left( (g^{TZT}_w+g^{YZY}_w)c_\ch-c^{(TZ)} s_\th s_\ch \right)\right)\right.	$	\\	
$		            $	&       $			$	&       $	  \left. - m_w s_\et \left(  2 (g^{XZZ}_w-g^{TXT}_w) s_\th + 2 c_\th \left(   (g^{TYT}_w-  g^{YZZ}_w)c_\ch+c^{(TY)}_w s_\th s_\ch  \right)  + c^{(TX)} c_\th^2 s_{2\ch} \right)\right]/m_w^2	$	\\	
$    A_{cS}^{(c,1)}	  $	&	$	\bE		$	&	$	\left[2 (g_w^{TZT}+g_w^{XZX}) m_w s_\th-\left(b_w^{T}+(g_w^{XZY}+g_w^{YZX}-g_w^{XYZ}) m_w\right) c_\th c_\chi+c_w^{(TZ)} m_w c^2_\th s_{2\ch}\right]/m_w^2		$	\\	
$    A_{sC}^{(c,1)}	  $	&	$	\bE		$	&	$	\left[ c_\et \left(  \left( b^T_w -m_w(g^{XYZ}_w+g^{XZY}_w+g^{YZX}_w) \right) c_\th c_\ch- 2 m_w (g^{TZT}_w+g^{YZY})s_\th  -c^{(TZ)} m_w c_\th^2 s_{2\ch}   \right)\right.	$	\\	
$		            $	&	$			$	&	$	  \left. +s_\et m_w\left( 2 (g^{TXT}_w-g^{XZZ}_w)c_\th c_\ch+2 s_\th (g^{TYT}_w-g^{YZZ}_w+c^{(TX)}_w c_\th s_\ch )- c^{(TY)} c_\th^2 s_{2\ch} \right) \right]/m^2_w	$	\\	
$   A_{sS} ^{(c,1)}	  $	&	$	\bE		$	&	$	\left[\left(b_w^{T}+(g_w^{XZY}-g_w^{XYZ}+g_w^{YZX}) m_w\right) s_\th+2m_w c_\th \left((g_w^{TZT}+g_w^{XZX}) c_\ch-c_w^{(TZ)} s_\th s_\ch\right)\right]/m_w^2		$	\\	
$   A_{c2C}^{(c,1)}      $	&	$	\bE		$	&	$	c_\et \left[c_w^{(TX)} c_\chi s_{2\theta}+c_w^{(TY)}( c^2_\chi s^2_\theta+s^2_\chi-c^2_\theta)\right]/ m_w		$	\\	
$   A_{c2S}^{(c,1)}      $	&	$	\bE		$	&	$	-\left[c_w^{(TY)} c_\ch s_{2\th}-c_w^{(TX)}( c^2_\ch s^2_\th+s^2_\chi- c^2_\th)\right]/ m_w		$	\\	
$   A_{s2C}^{(c,1)}      $	&	$	\bE		$	&	$	c_\et \left[ c_w^{(TX)}(3 c_{2\th}+2c^2_\th c_{2\chi} -1)+4 c_w^{(TY)} c_\chi s_{2\th}\right]/4 m_w		$	\\	
$   A_{s2S}^{(c,1)}      $	&	$	\bE		$	&	$	-\left[c_w^{(TY)}(3 c_{2\th}+2 c^2_\theta c_{2\chi}-1)-4 c_w^{(TX)}c_\chi s_{2\theta}\right]/4 m_w			$	\\						
\hline
\hline
\end{tabular}
\end{table*}

The structure of the frequency shifts \rf{dwc1} predicts sidereal variation of $\om_c^w$ up to the second harmonic of the sidereal frequency in contrast to the case of $\de \om_{c,0}^w$, see Eq.~\eqref{dwc0}, 
that contains contributions up to the third harmonic of the sidereal frequency \cite{dr20},
as expected, since contributions from the nonminimal terms to $\de \om_{c}^w$ are disregarded. If these nonminimal terms are included, the sidereal variation would contain contributions up to the fourth harmonic of the sidereal frequency. Emerging at linear order in $\be$ is a variation of $\om_c^w$ with the first harmonic of the annual frequency $\Om_\oplus$. The other variations of $\om_c^w$ are the products between the first harmonics of $\Om_\oplus$ with the first and second harmonics of $\om_\oplus$.

We can repeat the approach for the anomaly frequency shifts
in \eq{ca-lv-min} in a similar way.  
At linear order in $\be$,
the anomaly frequency shifts due to Lorentz violation in terms of the Sun-centered frame SME coefficients can be expressed as
\beq
\de \om_a^w\simeq \de \om_{a,0}^w + \de \om_{a,1}^w,
\eeq
where $\de \om_{a,0}^w$, the shifts at the zeroth order in the boost, are described in detail in Ref.~\cite{dk16} 
and take the form specified in Eq.~\eqref{dwa0} in 
Appendix \ref{appA}.
The expressions of $\de \om_{a,1}^w$ can be decomposed into
\bea
\de \om_{a,1}^w
&=&
A_0^{(a,1)}+A_c^{(a,1)} \cos\om_\oplus T_\oplus+A_s^{(a,1)} \sin\om_\oplus T_\oplus
\nn\\
&&
+A_C^{(a,1)} \cos\Om_\oplus T+A_S^{(a,1)} \sin\Om_\oplus T
\nn\\
&&
+\cos\om_\oplus T_\oplus \left( A_{cC}^{(a,1)}  \cos \Om_\oplus T 
+A_{cS}^{(a,1)} \sin \Om_\oplus T \right)
\nn\\
&&
+\sin \om_\oplus T_\oplus \left( A_{sC}^{(a,1)}  \cos \Om_\oplus T 
+A_{sS}^{(a,1)} \sin \Om_\oplus T \right)
\nn\\
&&
+A_{c2}^{(a,1)} \cos 2\om_\oplus T_\oplus+A_{s2} \sin 2\om_\oplus T_\oplus ,
\label{dwa1}
\eea
where a similar notation $A_{\ast}^{(a,1)}$ is used for the amplitudes 
for the case of the anomaly frequency shifts and their expressions 
are given in Table \ref{aV} and Table \ref{aH} for a vertical and horizontal magnetic field,
respectively. 
The structure of the tables is the same as the ones described before for the cyclotron frequency shifts.

\renewcommand{\arraystretch}{1.5}
\begin{table*}
\caption{
\label{aV}
SCF expressions of anomaly frequency shifts for a vertical $\mbf B$.}
\setlength{\tabcolsep}{5pt}
\begin{tabular}{ccc}
\hline
\hline																							
Amplitude		    &		Boost 			&		Coefficient 			\\	
	 		        &		factor			&		combination		 	\\	\hline
$	A_0^{(a,1)}      $&	$	\beL		$	&	$	-\left(2 H_w^{TZ}-(d_w^{XY}-d_w^{YX}+2 g_w^{TZT}+g_w^{XZX}+g_w^{YZY}) m_w\right) s^2_\chi		$	\\	
$	A_{c}^{(a,1)}    $&	$	\beL		$	&	$	\left(H_w^{TX}+(d_w^{ZY}-g_w^{TXT}+g_w^{XYY}) m_w\right) s_{2\chi}		$	\\	
$	A_{s}^{(a,1)}    $&	$	\beL		$	&	$	\left(H_w^{TY}-(d_w^{ZX}+g_w^{TYT}+g_w^{XYX}) m_w\right) s_{2\chi}		$	\\	
$	A_{C}^{(a,1)}    $&	$	\bE		$	&	$	-2 c_\chi \left[\left(H_w^{TX}+(d_w^{ZY}-g_w^{TXT}+g_w^{XYY}) m_w\right) c_\eta-\left(b_w^{T}-(d_w^{TT}+d_w^{ZZ}+g_w^{XYZ}) m_w\right) s_\eta\right]		$	\\	
$	A_{S}^{(a,1)}    $&	$	\bE		$	&	$	-2 \left(H_w^{TY}-(d_w^{ZX}+g_w^{TYT}+g_w^{XYX}) m_w\right) c_\chi		$	\\	
$	A_{c2}^{(a,1)}   $&	$	\beL		$	&	$	(d_w^{XY}+d_w^{YX}-g_w^{XZX}+g_w^{YZY}) m_w s^2_\chi		$	\\	
$	A_{s2}^{(a,1)}   $&  	$	\beL		$	&	$	-(d_w^{XX}-d_w^{YY}+g_w^{XZY}+g_w^{YZX}) m_w s^2_\chi		$	\\	
$	A_{cC}^{(a,1)}   $&	$	\bE		$	&	$	2 \left[\left(H_w^{TZ}-(d_w^{XY}+g_w^{TZT}+g_w^{YZY}) m_w\right) c_\et-\left(H_w^{TY}+(d_w^{XZ}-g_w^{TYT}+g_w^{YZZ}) m_w\right) s_\et\right] s_\chi		$	\\	
$	A_{cS}^{(a,1)}   $&	$	\bE		$	&	$	-2\left(b_w^{T}-(d_w^{TT}+d_w^{XX}+g_w^{YZX}) m_w\right) s_\chi		$	\\	
$	A_{sC}^{(a,1)}   $&	$	\bE		$	&	$	2 \left[\left(b_w^{T}-(d_w^{TT}+d_w^{YY}-g_w^{XZY}) m_w\right) c_\eta+\left(H_w^{TX}-(d_w^{YZ}+g_w^{TXT}-g_w^{XZZ}) m_w\right) s_\et\right] s_\chi		$	\\	
$	A_{sS}^{(a,1)}   $&	$	\bE		$	&	$	2 \left(H_w^{TZ}-(-d_w^{YX}+g_w^{TZT}+g_w^{XZX}) m_w\right) s_\chi		$	\\	
\hline
\hline
\end{tabular}
\end{table*} 

\renewcommand{\arraystretch}{1.5}
\begin{table*}
\caption{
\label{aH}
SCF expressions of anomaly frequency shifts for a horizontal $\mbf B$.}
\setlength{\tabcolsep}{5pt}
\begin{tabular}{ccc}
\hline
\hline																							
Amplitude		    &		Boost 			&		Coefficient 			\\	
	 		        &		factor			&		combination		 	\\	\hline
$	A_0^{(a,1)}      $&	$	\beL		$	&	$	 -\left[\left(2 H_w^{TZ}-(d_w^{XY}-d_w^{YX}+2 g_w^{TZT}+g_w^{XZX}+g_w^{YZY}) m_w\right) c_\theta c_\chi	\right.	$	\\	
$		           $&	$			    $	&	$	  \left.+\left(2 b_w^{T}-(2 d_w^{TT}+d_w^{XX}+d_w^{YY}-g_w^{XZY}+g_w^{YZX}) m_w\right) s_\th\right] s_\chi		$	\\	
$	A_{c}^{(a,1)}    $&	$	\beL		$	&	$	-2\left (H_w^{TX}+(d_w^{ZY}-g_w^{TXT}+g_w^{XYY}) m_w\right) c_\theta s^2_\chi		$	\\	
$	A_{s}^{(a,1)}    $&	$	\beL		$	&	$	-2 \left(H_w^{TY}-(d_w^{ZX}+g_w^{TYT}+g_w^{XYX}) m_w\right) c_\theta s^2_\chi		$	\\	
$	A_{C}^{(a,1)}    $&	$	\bE		$	&	$	2 c_\th \left[\left(H_w^{TX}+(d_w^{ZY}-g_w^{TXT}+g_w^{XYY}) m_w\right) c_\eta-\left(b_w^{T}-(d_w^{TT}+d_w^{ZZ}+g_w^{XYZ}) m_w\right) s_\et\right] s_\chi		$	\\	
$	A_{S}^{(a,1)}    $&	$	\bE		$	&	$	2 \left(H_w^{TY}-(d_w^{ZX}+g_w^{TYT}+g_w^{XYX}) m_w\right) c_\theta s_\chi		$	\\	
$	A_{c2}^{(a,1)}   $&	$	\beL		$	&	$	m_w \left((d_w^{XY}+d_w^{YX}-g_w^{XZX}+g_w^{YZY}) c_\theta c_\chi-(d_w^{XX}-d_w^{YY}+g_w^{XZY}+g_w^{YZX}) s_\th\right) s_\chi		$	\\	
$	A_{s2}^{(a,1)}   $&	$	\beL		$	&	$	-m_w \left((d_w^{XX}-d_w^{YY}+g_w^{XZY}+g_w^{YZX}) c_\theta c_\chi+(d_w^{XY}+d_w^{YX}-g_w^{XZX}+g_w^{YZY}) s_\theta\right) s_\chi		$	\\	
$	A_{cC}^{(a,1)}   $&	$	\bE		$	&	$	2c_\eta \left[\left(H_w^{TZ}-(d_w^{XY}+g_w^{TZT}+g_w^{YZY}) m_w\left) c_\th c_\chi+\right(b_w^{T}-(d_w^{TT}+d_w^{YY}-g_w^{XZY}) m_w\right) s_\th\right]$	\\	
$		           $&	$			    $	&	$	 +2s_\eta \left[\left(H_w^{TX}-(d_w^{YZ}+g_w^{TXT}-g_w^{XZZ}) m_w\right) s_\th-\left(H_w^{TY}+(d_w^{XZ}-g_w^{TYT}+g_w^{YZZ}) m_w\right) c_\th c_\chi\right]		$	\\	
$	A_{cS}^{(a,1)}   $&	$	\bE		$	&	$	-2 \left(b_w^{T}-(d_w^{TT}+d_w^{XX}+g_w^{YZX}) m_w\right) c_\th c_\chi+2 \left(H_w^{TZ}+(d_w^{YX}-g_w^{TZT}-g_w^{XZX}) m_w\right) s_\theta 	$	\\	
$	A_{sC}^{(a,1)}   $&	$	\bE		$	&	$	2 c_\eta \left[\left(b_w^{T}-(d_w^{TT}+d_w^{YY}-g_w^{XZY}) m_w\right) c_\th c_\chi-\left(H_w^{TZ}-(d_w^{XY}+g_w^{TZT}+g_w^{YZY}) m_w\right) s_\th\right]		$	\\	
$		           $&	$			    $	&	$	  +2 s_\eta \left[\left(H_w^{TX}-(d_w^{YZ}+g_w^{TXT}-g_w^{XZZ}) m_w\right) c_\th c_\chi+\left(H_w^{TY}+(d_w^{XZ}-g_w^{TYT}+g_w^{YZZ}) m_w\right) s_\theta\right]		$	\\	
$	A_{sS}^{(a,1)}   $&	$	\bE		$	&	$	2 \left[\left(H_w^{TZ}+(d_w^{YX}-g_w^{TZT}-g_w^{XZX}) m_w\right) c_\th  c_\chi+\left(b_w^{T}-(d_w^{TT}+d_w^{XX}+g_w^{YZX}) m_w\right) s_\theta\right]		$	\\	
\hline
\hline
\end{tabular}
\end{table*} 

The signals for Lorentz violation predicted by Eq.~\eqref{dwa1} include a sidereal variation of $\om_a^w$ with the first and second harmonic of the sidereal frequency similar to the signals predicted at the zeroth order in $\be$. Introduced at linear order in $\be$ is a variation of $\om_a$ with the first harmonic of the annual frequency $\Om_\oplus$ and the product between the first harmonics of $\Om_\oplus$ with the first harmonic of $\om_\oplus$.

The frequency shifts $\de \om_{c}^{\ol w}$ and $\de \om_{a}^{\ol w}$   
for antifermions, 
see \eq{cas-lv-min}, at linear order in the boost are given by Eqs.~\eqref{dwc1} and \eqref{dwa1} with some modifications.   
The amplitudes $A_{\ast}^{(c,1)}$ and $A_{\ast}^{(a,1)}$ are replaced by ${\ol A}^{(c,1)}_{\ast}$ and ${\ol A}^{(a,1)}_{\ast}$, 
in which that the signs
 in front of the CPT-odd SME coefficients $b^{\mu}_w$ and $g^{\mu \nu\al}_w$ are reversed.  As an example, the expression for ${\ol A}_{c}^{(a,1)} $ is 
\beq
{\ol A}_{c}^{(a,1)}  =\beL \sin^2\chi (H_w^{TX}+\left(d_w^{ZY}+g_w^{TXT}-g_w^{XYY}) m_w\right), 	
\eeq
compared to
\beq
{ A}_{c}^{(a,1)}  =\beL \sin^2\chi (H_w^{TX}+\left(d_w^{ZY}-g_w^{TXT}+g_w^{XYY}) m_w\right). 	
\eeq

Before concluding this section it is convenient to introduce some terminology to facilitate the discussion of the signals for Lorentz violation. We define the pure sidereal variation of the cyclotron frequency shifts at linear order in $\be$ by
\bea
\fr{(\de \om_{c,1}^w)_{\rm sid}}{eB}&=&A_c^{(c,1)} \cos\om_\oplus T_\oplus+A_s^{(c,1)} \sin\om_\oplus T_\oplus\nn \\
&&+A_{c2}^{(c,1)} \cos 2 \om_\oplus T_\oplus+A_{s2}^{(c,1)} \sin 2 \om_\oplus T_\oplus ,
\nn\\
\eea
and for the anomaly frequency shifts, we define it by
\bea
 (\de\om_{a,1}^w)_{\rm sid}
 &=&A_c^{(a,1)} \cos\om_\oplus T_\oplus+A_s^{(a,1)} \sin\om_\oplus T_\oplus
 \nn\\
 &&
+A_{c2}^{(a,1)} \cos 2\om_\oplus T_\oplus+A_{s2} \sin 2\om_\oplus T_\oplus .
\nn\\
\eea
The pure annual variations are defined by 
\beq
\fr{(\de \om_{c,1}^w)_{\rm ann}}{eB}=A_C^{(c,1)} \cos\Om_\oplus T+A_S^{(c,1)} \sin\Om_\oplus T
\eeq
for the cyclotron frequency shifts, and by
\beq
(\de \om_{a,1}^w)_{\rm ann}=A_C^{(a,1)} \cos\Om_\oplus T+A_S^{(a,1)} \sin\Om_\oplus T
\eeq
for the anomaly frequency shifts.  
Finally, we define the mixed annual-sidereal variation of the cyclotron and anomaly frequency shifts by
\bea
&&\fr{(\de \om_{c,1}^w)_{\rm mix}}{eB}
\nn\\
=&&
\cos \om_\oplus T_\oplus \left(A_{cC}^{(c,1)} \cos\Om_\oplus T+A_{cS}^{(c,1)} \sin\Om_\oplus T\right)
\nn\\
&&
+\sin\om_\oplus T_\oplus \left( A_{sC}^{(c,1)} \cos\Om_\oplus T+A_{sS}^{(c,1)} \sin\Om_\oplus T\right)
\nn\\
&&
+\cos 2\om_\oplus T_\oplus \left(A_{c2C}^{(c,1)}\cos \Om_\oplus T+A_{c2S}^{(c,1)}\sin\Om_\oplus T\right)
\nn\\
&&
+\sin 2\om_\oplus T_\oplus\left(A_{s2C}^{(c,1)} \cos \Om_\oplus T+A_{s2S}^{(c,1)}\sin\Om_\oplus T \right)
\nn\\
\eea
and 
\bea
&&(\de \om_{a,1}^w)_{\rm mix} 
\nn\\
=&&
\cos \om_\oplus T_\oplus \left(A_{cC}^{(a,1)}\cos\Om_\oplus T+A_{cS}^{(a,1)} \sin\Om_\oplus T\right)
\nn\\
&&
+\sin \om_\oplus T_\oplus \left(A_{sC}^{(a,1)}\cos\Om_\oplus T+A_{sS}^{(a,1)} \sin\Om_\oplus T\right).
\nn\\
\eea
We use the same terminology for the frequency shifts for antiparticles by replacing the amplitudes 
$A_{\ast}^{(c,1)}$, $A_{\ast}^{(a,1)}$ by ${\ol A}_{\ast}^{(c,1)}$, ${\ol A}_{\ast}^{(a,1)}$.

\section{Experiments}
\label{experiments}

In this section, 
we analyze several Penning-trap experiments 
that measure the charge-to-mass ratios,
the $g$ factors,
and their comparisons between particles and antiparticles,  
and use the reported experimental measurements to constrain the relevant 
Sun-centered frame SME coefficients  
that are associated to linear boost corrections.
The experiments chosen for discussion here are listed in 
\Tab{exp-p}.
For each experiment, 
we include the relevant particle species,
the colatitude $\ch$ of the laboratory,
the direction and magnitude of the magnetic field $\mbf B$ used in the trap
(for a horizontal magnetic field, the angle in the parenthesis specifies its
direction in the horizontal plane,
measured from the local south in the counterclockwise direction),
and the translated precision from the reported measurement
by each experiment in terms of the upper limit
of the relevant frequencies in units of GeV 
(details are given in the next two subsections). 
We note that the Penning-trap experiments 
involving electrons and positrons are not included in 
\Tab{exp-p} 
as the limits of the electron coefficients for Lorentz and CPT violation 
obtained from these experiments are $10^{7}-10^{16}$ orders of magnitude larger than 
the current best bounds obtained from other experimental systems.
For example, 
the limit on the magnitude of the electron coefficient $b_e^T$ 
obtained in this work is $|b_e^T| \lsim 10^{-11}$ GeV,
while experiments using a torsion pendulum 
have constrained this coefficient to a $10^{-27}$ GeV level
\cite{torsion}. 
Therefore, 
we restrict our attention to Penning-trap experiments involving protons and antiprotons in this work.
 
\renewcommand{\arraystretch}{1.5}
\begin{table*}
\caption{
\label{exp-p}
Experimental quantities for relevant Penning-trap experiments.}
\setlength{\tabcolsep}{5pt}
\begin{tabular}{cccccc}
\hline
\hline																						
																			
Experiment	&	Species	&	$	\chi	$	&	$\mbf B$ direction	&	$B$	&	 	Precision			 		\\	\hline
ATRAP \cite{ga99}	&	$p$, $\ol p$	&	$	43.8^{\circ}	$	&	upward	&	5.85 T	&	$	|\delta \omega_c^{\ol p}-1.001\delta \omega_c^{p}|_\text {const } 	<	3.33 \times 10^{-26}	$	GeV	\\	
BASE \cite{ul15}	&	$p$, $\ol p$	&	$	43.8^{\circ}	$	&	horizontal  ($120^{\circ}$)	&	1.946 T	&	$	|\delta \omega_c^{\ol p}-1.001\delta \omega_c^{p}|_\text {const } 	<	8.46 \times 10^{-27}	$	GeV	\\	
	&		&	 		 	&		&		&	$	|\delta \omega_c^{\ol p}-1.001\delta \omega_c^{p}|_\text {1st } 	<	8.83 \times 10^{-26}	$	GeV	\\	
BASE \cite{19sm}	&	$\ol p$	&	$	43.8^{\circ}	$	&	horizontal  ($120^{\circ}$)	&	1.946 T	&	$	|\delta \omega_a^{\ol p}|_\text {1st } 	<	1.81 \times 10^{-24}	$	GeV	\\	
	&		&	 		 	&		&		&	$	|\delta \omega_a^{\ol p}|_\text {2nd } 	<	1.81 \times 10^{-24}	$	GeV	\\	
BASE \cite{17sc, 17sm}	&	$p$, $\ol p$	&	$	40.0^{\circ}, 43.8^{\circ}	$	&	horizontal ($18^{\circ}, 120^{\circ}$)	&	1.9 T	&	$	|\delta \omega_a^{p}-0.98\delta \omega_a^{\ol p}|_\text {const } 	<	9.53 \times 10^{-25}	$	GeV	\\		
\hline
\hline
\end{tabular}
\end{table*}

\subsection{The charge-to-mass ratios}
\label{the charge-to-mass ratios}

It is evident from \eq{qm} that the charge-to-mass ratio of a charged particle or antiparticle 
confined in a Penning trap is related to the ratios of its cyclotron frequency
and the magnetic field used in the trap.
In the presence of Lorentz and CPT violation, 
corrections can be introduced to the cyclotron frequencies,
as shown by expressions \rf{ca-lv-min} and \rf{cas-lv-min}
in the context of the minimal SME.
These corrections are controlled by a set of tilde effective coefficients  
$\bptw w {3}$, 
$\ctw w {11} + \ctw w {22}$,
$\bptws w {3}$, 
and
$\ctws w {11} + \ctws w {22}$
in the apparatus frame.
The fundamental coefficients for Lorentz violation in these tilde effective coefficients are given
by definitions \rf{tild-min} and \rf{tilds-min}.
The coefficients for Lorentz violation appearing in these expressions have nontrivial 
transformation properties under rotations and boosts to the Sun-centered frame,
thus introducing time-varying signals to the measured cyclotron frequencies
given by \eq{dwc1},
as discussed in detail in 
\Sec{transformations}.
Performing a time-variation analysis of the measurement data 
can extract the time dependence of the cyclotron frequencies and
set bounds on relevant coefficients for Lorentz violation. 
 
Results \rf{ca-lv-min} and \rf{cas-lv-min} also show that 
the cyclotron frequency shifts due to Lorentz and CPT violation 
for a particle are different from these for its corresponding antiparticle,
due to the sign changes of all the CPT-odd coefficients 
in these two expressions.
For experiments comparing the charge-to-mass ratios between
a particle and its corresponding antiparticle,
the difference in the charge-to-mass ratios corresponds to   
that in the cyclotron frequency shifts, 
\bea
\label{ratio-lv}
\dfrac{(|q|/m)_{\ol{w}}}{(|q|/m)_{w}} - 1
\longleftrightarrow
\dfrac{\om_c^{\ol{w}}}{\om_c^{w}} - 1
=
\dfrac{\de \om_c^{\ol{w}} -  \de \om_c^{w}} {\om_c^{w}},
\eea
where the Lorentz- and CPT-invariant pieces 
in the cyclotron frequencies are exactly canceled
by the CPT theorem if the same magnetic field is used.
The notation~$\longleftrightarrow$ indicates the correspondence between 
the experimental interpreted charge-to-mass ratio comparison
and the measured frequency difference.
Based on relation~\rf{ratio-lv},
the measurement precision in the difference
$ |q|/m)_{\ol{w}}/(|q|/m)_{w} - 1$
reported by an experiment can be used 
to extract limits on the relevant coefficients for Lorentz violation
that appear in $\de \omega_c^{w}$ and $\de \omega_c^{\ol{w}}$. 

Before we start the analysis of the experiment results to extract the limits
on the coefficients for Lorentz violation,
we want to point out a subtlety related to the particle species used in the experiments.  
For the experiment comparing the charge-to-mass ratios between protons and antiprotons,
most experiments use a hydrogen ion (H$^-$) as a proxy for the proton 
to eliminate systematic shifts caused by polarity switching of the trapping voltages.
This modifies relation~\rf{ratio-lv} to
\beq
\label{ratioH1}
\dfrac{(|q|/m)_{\bar{p}}}{(|q|/m)_{p}} - 1
=
\dfrac{(|q|/m)_{\bar{p}}}{R (|q|/m)_{\rm{H}^-}} - 1
\longleftrightarrow
\dfrac{\de \om_c^{\bar{p}} - R \de \om_c^{\rm{H}^-}} {R \om_c^{\rm{H}^-}} ,
\eeq
where $R=m_{\rm{H}^-}/m_p = 1.001089218754$
is the ratio of the mass between a hydrogen ion and a 
proton~\cite{ul15},
$\om_c^{\rm{H}^-}$ is the cyclotron
frequency for the hydrogen ion,
and $\de \om_c^{\rm{H}^-}$ is its corresponding shift. 
To obtain $\de \om_c^{\rm{H}^-}$,
one can apply $w=\rm{H}^-$ in expression~\rf{ca-lv}
and the related tilde coefficients for Lorentz violation 
become the effective ones for hydrogen ions.   
Expressing these effective coefficients in terms of the 
corresponding fundamental coefficients for the hydrogen ion constituents,
the electron and proton coefficients,
is challenging due to nonperturbative issues including binding effects 
in the composite hydrogen ion. 
However, 
an approximation to these coefficient relations can be obtained by 
treating the wave function of the hydrogen ion as a product of 
the wave functions of a proton and two electrons.
Applying perturbation theory at the lowest order
and ignoring the related binding energies,
the cyclotron frequency shifts $\de \om_c^{\rm{H}^-}$ of the hydrogen ion 
due to Lorentz and CPT violation can then be approximated as the sum of these for its constituents,
$\de \om_{c}^{\rm{H}^-} \simeq \de \om_{c}^{p} + 2 \de \om_{c}^{e^-}$.
The term $\de \om_{c}^{e^-}$ contains coefficients for Lorentz violation in the electron sector.
Compared to the shifts in the proton cyclotron frequencies,   
the ones arising from the electron cyclotron frequency shifts are suppressed
by a factor of $m_e/m_p\simeq 10^{-3}$.
For this reason,
we can ignore the term $2\de \om_{c}^{e^-}$ in the shifts
of the hydrogen ion cyclotron frequency shifts and keep contributions 
from the protons only. 
Under this assumption,
relation~\rf{ratioH1} now becomes
\bea
\label{ratioH2}
\dfrac{(|q|/m)_{\bar{p}}}{(|q|/m)_{p}} - 1
\longleftrightarrow
\dfrac{\de \om_c^{\bar{p}} - R \de \om_c^{p} } {R \om_c^{\rm{H}^-}} .
\eea
In the following two subsections, 
we will use relation~\rf{ratioH2},
together with the reported precisions 
of the charge-to-mass ratio comparisons between protons and antiprotons
and appropriate numerical values of $\om_c^{\rm{H}^-}$ for different Penning-trap
experiments to constrain the relevant coefficients for Lorentz violation.

\subsubsection{ATRAP at CERN}
\label{ATRAP at CERN}

In a Penning-trap experiment located at CERN by the ATRAP collaboration,
Gabrielse and his group achieved a precision of 90 ppt 
for the difference 
of the proton-antiproton charge-to-mass ratio 
comparison~\cite{ga99}. 
The experiment applied an upward uniform magnetic field
$B = 5.85$~T
 in the trap.
The reported result was obtained by taking the time average of the cyclotron frequency measurements.  
This procedure resulted in a suppression of the time-dependent terms 
and implies that only the coefficients that appear in the 
constant term in the time variations of the cyclotron frequencies 
can be constrained using the experimental result.  

Applying expression \rf{ratioH2} by taking the reported precision of 90 ppt for
$(|q|/m)_{\bar{p}}/(|q|/m)_{p} - 1$
and identifying $\om_c^{\rm{H}^-} =  2\pi \times 89.3$ MHz 
given in the ATRAP experiment,
the following limit can be obtained,
\beq
\label{cm-limit-atrap}
| \de \om_c^{\bar{p}} - 1.001 \de \om_c^{p} |_{\rm{const}} 
\lsim 3.33 \times 10^{-26}\ \rm{GeV},
\eeq
where the subscript ``const" indicates that only the constant terms 
contributing to the frequency shifts are relevant to the above limit.

\subsubsection{BASE at CERN}
\label{BASE at CERN}

More recently, 
another Penning-trap experiment at CERN by the BASE collaboration
led by Ulmer improved the comparison to a sensitivity of 69 
ppt~\cite{ul15},
by applying a horizontal magnetic field $B = 1.946$ T 
pointing $\th=120^\circ$ from the local south in the counterclockwise direction.
The BASE experiment analyzed the data of the charge-to-mass ratio comparisons
to search for both time-averaged effects and 
sidereal variations in the first harmonic $\om_\oplus$ of the Earth's rotation frequency.
Focusing on the linear order boost effects,
the reported results can be taken to set bounds on not only the constant terms
$A_0^{(c,1)}$ and ${\ol A}_0^{(c,1)}$, 
but also on the amplitudes 
$A_c^{(c,1)}$, $A_s^{(c,1)}$, ${\ol A}_c^{(c,1)}$ and ${\ol A}_s^{(c,1)}$,
which are proportional to the first harmonic 
of the variations in the sidereal frequencies.  

Using the reported 69 ppt for the time-averaged precision
and 720 ppt for the limit of the first harmonic amplitude for the comparison \rf{ratioH2}
and taking $\om_c^{\rm{H}^-} =  2\pi \times 29.6$ MHz for the BASE experiment,
the following limits are obtained,
\beq
\label{cm-limit-base-cst}
| \de \om_c^{\bar{p}} - 1.001 \de \om_c^{p} |_{\rm{const}} 
\lsim 8.46 \times 10^{-27}\ \rm{GeV}
\eeq
and 
\beq
\label{cm-limit-base-1st}
| \de \om_c^{\bar{p}} - 1.001 \de \om_c^{p} |_{\rm{1st}} 
\lsim 8.83 \times 10^{-26}\ \rm{GeV},
\eeq
where the subscript ``const" in the limit \rf{cm-limit-base-cst}
takes the same meaning as the one in \rf{cm-limit-atrap},
while the subscript ``1st" in the limit \rf{cm-limit-base-1st} specifies
the amplitude of the first harmonic in the sidereal variation.

\subsection{The $g$ factors}
\label{the g factors}

Another intrinsic fundamental quantity of a particle is known as the $g$ factor, 
which is associated to the particle's anomalous magnetic moment.
The $g$ factor of a particle 
can be determined using a Penning trap 
by measuring the ratio of its anomaly frequency 
and cyclotron frequency,
as shown by expression~\rf{g}.
Similar to the discussion of the charge-to-mass ratios in
Subsection~\ref{the charge-to-mass ratios},
Lorentz and CPT violation can introduce shifts 
to both cyclotron and anomaly frequencies of the particle in the trap
according to expressions \rf{ca-lv-min} and \rf{cas-lv-min}.
Compared to the shifts in the anomaly frequencies 
$\de \om_a^w$ and $\de {\om_a^{\ol {w}}}$ , 
contributions to the cyclotron frequencies 
$\de \om_c^w$ and $\de {\om_c^{\ol {w}}}$ 
are suppressed by factors of $eB/m_w^2$.
Even for a comparatively large magnetic field of $B \simeq 5$ T in a Penning trap,
these factors are at orders of $eB/m_p^2 \simeq 10^{-16}$ for protons and antiprotons.
Therefore,
to obtain the dominant effects due to Lorentz and CPT violation, 
we can ignore the shifts in the cyclotron frequencies 
and focus only on these in the anomaly frequencies.
According to expressions \rf{ca-lv-min} and \rf{cas-lv-min}, 
the shifts in the anomaly frequencies are controlled by 
the tilde combinations 
$\btw w 3$ and $\btws w 3$
in the apparatus frame,
with their definitions in terms of the fundamental coefficients for Lorentz violation
given by 
expressions \rf{tild-min} and \rf{tilds-min}.
The rotation and boost transformations of these fundamental coefficients for Lorentz violation 
from the apparatus frame to the Sun-centered frame
introduce time-varying signals in the measurements of the $g$ factors.
A time-variation analysis of the anomaly frequency measurements would 
permit constraints on the relevant coefficients for Lorentz violation. 

Since the shifts in the anomaly frequencies due to Lorentz and CPT violation
between a particle and an antiparticle are different,
as shown in results \rf{ca-lv-min} and \rf{cas-lv-min},
for experiments comparing the $g$ factors between a particle and its antiparticle,
the comparison is related to the difference in the anomaly frequencies,
given by
\bea
\label{gratio-lv}
\half (g_w - g_{\ol w})
\longleftrightarrow
\fr {\om_a^w}{\om_c^w} - \fr {\om_a^{\ol w}}{\om_c^{\ol w}} =
\fr {\de\om_a^w}{\om_c^w} - \fr {\de\om_a^{\ol w}} {\om_c^{\ol w}},
\eea 
where again all Lorentz- and CPT-invariant contributions are canceled out 
on the righthand side. 
Note relation \rf{gratio-lv} doesn't require the use of the same magnetic field 
to measure the $g$ factors of a particle and an antiparticle. 
If different magnetic fields are used in the traps, 
$\om_c^{w}$ and $\om_c^{\ol w}$ would have different values, 
and the coefficients in $\de\om_a^w$ and $\de\om_a^{\ol w}$
would have different transformation expressions as they depend 
on the field orientations used in the traps. 
In the next two subsections, 
we will analyze Penning-trap experiments that measure the $g$ factors of 
protons, antiprotons, and their comparisons,
and use the reported precisions for these measurements 
to constrain the relevant coefficients for Lorentz violation in the proton sector.

\subsubsection{BASE at CERN}
\label{BASE at CERN}

The measurement of the $g$ factor for antiprotons 
has reached a record precision of 1.5 ppb by the BASE collaboration 
using a Penning trap located at CERN,
with a horizontal magnetic field $B=1.946$ T 
at $\th=120^\circ$ from the local south 
~\cite{17sm}.
A sidereal-variation analysis of the Larmor frequencies was performed
at the frequencies of $\om_{\oplus}$ and $2\om_{\oplus}$,
yielding a precision of 5.3 ppb and 5.2 ppb,
respectively. 
Since $\om_L=\om_a+\om_c$,
any shift in the Larmor frequency of the antiproton is the sum of the shifts in its 
anomaly frequency and the cyclotron frequency.
However, 
as discussed at the beginning of this subsection, 
shifts in the cyclotron frequencies are suppressed by these 
in the anomaly frequencies by a factor of $eB/m_p^2 \simeq 10^{-16}$.
Keeping only the dominant contributions to the Larmor frequency,
we have $\de \om_L=\de \om_a$. 
Identifying $\om_L^{\ol{p}}=2\pi \times 82.82$ MHz,
together with the reported precisions of 5.3 ppb and 5.2 ppb
of the sidereal variations in $\om_{\oplus}$ and $2\om_{\oplus}$,
we have the following limits in natural units
\beq
\label{g-limit-base-1st}
| \de \om_a^{\ol{p}}|_{\rm{1st}} 
\lsim 1.81 \times 10^{-24} \ \rm{GeV},
\eeq
and 
\beq
\label{g-limit-base-2nd}
| \de \om_a^{\ol{p}}|_{\rm{2nd}} 
\lsim 1.81 \times 10^{-24}\ \rm{GeV},
\eeq
where the subscripts ``1st" and ``2nd'' 
take the same meaning as before.

\subsubsection{BASE at Mainz and CERN}
\label{BASE at Mainz and CERN}

The proton's $g$ factor has been measured to 
a record precision of 0.3 ppb,
by the same BASE collaboration using a Penning trap 
located at Mainz
with a horizontal magnetic field $B=1.9$ T 
at $\th=18^\circ$ from the local south 
~\cite{17sc}.
At the end of this subsection, 
we combine the proton's $g$ factor measurement with that of an antiproton,
discussed in the preceding section,
to extract limits of additional coefficients for Lorentz violation in the proton sector. 
Combining the reported precisions of 0.3 ppb (proton) and 1.5 ppb (antiproton)
for the time-averaged measurements,
and identifying $\om_c^{p}=2\pi\times 28.96$ MHz
and 
$\om_c^{\ol p}=2\pi\times 29.66$ MHz
for each experiment,
comparison \rf{gratio-lv} gives 
\bea
\label{g-limit-basecom-cst}
| \de\om_a^p - 0.98 \de\om_a^{\ol p} |_{\rm{const}} 
\lsim  9.53 \times 10^{-25}\ \rm{GeV},
\eea
where the same subscript ``const" is used to specify 
only the constant terms in the transformation are relevant to the above limit. 
The factor 0.98 is the ratio of $w_c^p/w_c^{\ol p}$ 
due to the different cyclotron frequencies in the two experiments.

\subsection{Results}
\label{results}

The limits appearing in expressions
\rf{cm-limit-atrap},
\rf{cm-limit-base-cst},
\rf{cm-limit-base-1st},
\rf{g-limit-base-1st},
\rf{g-limit-base-2nd},
and
\rf{g-limit-basecom-cst}
contain the shifts in the cyclotron and anomaly frequencies 
of protons and antiprotons.
The exact expressions of these frequency shifts in terms of the fundamental coefficients 
for Lorentz violation in the Sun-centered frame depend,
in general, 
on the field configuration of each experiment,
as given by
\Tab{cV},
\Tab{cH},
\Tab{aV},
and 
\Tab{aH}
in 
\Sec{transformations}. 
Extracting the terms in the relevant amplitudes 
according to the subscripts of the limit expressions,
together with the corresponding experimental values
 listed in 
\Tab{exp-p} 
for each experiment,
the limits on the relevant coefficients for Lorentz violation can be obtained.  
To illustrate this idea in more detail,
we provide here an example that analyzes the BASE experiment comparing 
the charge-to-mass ratios between protons and antiprotons at CERN. 

Since the BASE experiment at CERN applied a magnetic field of $B=1.946$ T 
in the horizontal direction,
and limits  
\rf{cm-limit-base-cst} and \rf{cm-limit-base-1st}
include shifts in the cyclotron frequencies of protons and antiprotons,
the corresponding table that lists the transformation results is identified as
\Tab{cH}. 
The subscript ``const"  
 in limit \rf{cm-limit-base-cst} suggests that only the constant 
terms in the time variations can be constrained using this limit,
which are the terms appearing in the amplitudes 
$A_0^{(c,0)}$, $A_0^{(c,1)}$, ${\ol A}_0^{(c,0)}$, and ${\ol A}_0^{(c,1)}$
in the first expression in
\eq{wcSidAppCp}.
Similarly, 
the subscript ``1st" in limit \rf{cm-limit-base-1st} implies that the terms corresponding 
to both $\cos \om_\oplus T_\oplus$ and $\sin \om_\oplus T_\oplus$ can be bounded.
These terms can be identified as the ones in the amplitudes 
$A_c^{(c,0)}$, $A_s^{(c,0)}$, $A_c^{(c,1)}$, $A_s^{(c,1)}$, 
${\ol A}_c^{(c,0)}$, ${\ol A}_s^{(c,0)}$, ${\ol A}_c^{(c,1)}$, and ${\ol A}_s^{(s,1)}$
in the second expression in 
\eq{wcSidAppCp}.
The amplitudes with a 0 in the superscripts denote contributions at the zeroth boost order
and they have been studied in detail in Ref.~\cite{dr20},  
therefore, 
we focus here on the reinterpretation of limits  
\rf{cm-limit-base-cst} and \rf{cm-limit-base-1st}
on the amplitudes at the linear boost order.
Keeping only the amplitudes at linear boost order 
in limits \rf{cm-limit-base-cst} and \rf{cm-limit-base-1st}
with the laboratory colatitude $\chi=43.8^\circ$ listed in 
\Tab{exp-p}
and taking $\beL \simeq 1.6 \times 10^{-6}$,
 the limits on the relevant combination of coefficients for Lorentz violation
 are thus obtained.  
 We note that in deriving the limits from the 
BASE at Mainz and CERN experiments 
(discussed in Subsection 
\ref{BASE at Mainz and CERN}) 
using this method,
since the two BASE experiments used magnetic fields 
in different directions,
the angle $\th$ in the transformations of  $\de\om_a^p$ and  $\de\om_a^{\ol p}$
takes different values. 
We also note that 
Appendix~\ref{appA}
provides an extended discussion 
on combining the present work at linear boost order  
with the previous works on zeroth boost order  
\cite{dr20, dk16}.

Some intuition about the scope of the limits 
on the individual fundamental coefficients for Lorentz violation
appearing in the above limits can be obtained using 
a common practice adopted in many subfields searching for 
Lorentz and CPT 
violation~\cite{tables},
which assumes that only one individual coefficient is nonzero
at a time and neglects any possible cancellation among different coefficients.  
This procedure offers a reasonable insight into the maximum conceivable sensitivity
from the constraint to each individual coefficient,
allowing us to quickly compare the sensitivities of different experiments to the individual coefficients 
and to recognize which sectors of the SME remain poorly explored.  
Following this spirit, 
we set limits on individual components of the coefficients for Lorentz violation 
and list them in 
\Tab{limits}.
In the table, 
the first column lists the components of the coefficients 
for Lorentz violation in the Sun-centered frame,
with parentheses on $n$ indices implying symmetrization
and brackets on $n$ indices indicating antisymmetrization,
both with a factor of $1/n!$.
The second column displays the constraints obtained from other work. 
A blank space in this column indicates that we did not find any previous work 
that has imposed a limit on this coefficient in the literature. 
We also verified that the listed coefficient does not contribute to 
any of the constraints reported in the most recent edition 
of the Tables for Lorentz and CPT 
Violation~\cite{tables}, 
including the ones listed in tables D9 and D10 of this reference. 
The third column displays the constraints obtained by this work.
The final column gives the relevant experiments used to obtain the constraints,
in which the notations in the parentheses indicate the relevant 
particle species and quantities measured by the experiment. 
For example, 
($p$ $g$) means $g$ factor measurements for protons
and 
($p-\ol{p}$ $c/m$) implies proton-antiproton 
charge-to-mass ratio comparisons. 
As shown in \Tab{limits}, 
this work obtained 18 {\it first-time} limits on coefficients that have not been bounded before
and improved the constraints of 2 additional coefficients,
$b_p^T$ and $d_p^{TT}$ ,
by about 4 and 10 orders of magnitude
compared to the previous constraints. 
The limits on $c_p^{(TX)}$,  $c_p^{(TY)}$, and $c_p^{(TZ)}$
are not comparable to the previous ones, 
but for completeness,
we also include them in the table.

\renewcommand{\arraystretch}{1.5}
\begin{table*}
\caption{
\label{limits}
Limits on the proton coefficients in the minimal SME.}
\setlength{\tabcolsep}{7pt}
\begin{tabular}{clll}
\hline
\hline																																					Coefficient			&			Previous constraint				 	&			This work				&	Experiment	\\	\hline
$	|	b_p^{T}	|	$	&	$	3.8	\times	10^{-15}	$	GeV	\cite{2018Ferrari}	&	$	3.8	\times	10^{-19}	$	GeV	&	BASE at Mainz ($p$ $g$)	\\	
$	|	c_p^{(TX)}	|	$	&	$	1.0	\times	10^{-20}	$		\cite{2016LeBars}	&	$	5.1	\times	10^{-1}	$		&	BASE at CERN ($p-\ol{p}$ $c/m$)	\\	
$	|	c_p^{(TY)}	|	$	&	$	1.0	\times	10^{-20}	$		\cite{2016LeBars}	&	$	5.1	\times	10^{-1}	$		&	BASE at CERN ($p-\ol{p}$ $c/m$)	\\	
$	|	c_p^{(TZ)}	|	$	&	$	1.0	\times	10^{-20}	$		\cite{2016LeBars}	&	$	2.1	\times	10^{-1}	$		&	BASE at CERN ($p-\ol{p}$ $c/m$)	\\	
$	|	d_p^{TT}	|	$	&	$	3.0	\times	10^{-8}	$		\cite{sf15}	&	$	8.2	\times	10^{-19}	$		&	BASE at Mainz ($p$ $g$)	\\	
$	|	d_p^{XX}	|	$	&								&	$	1.6	\times	10^{-18}	$		&	BASE at Mainz ($p$ $g$)	\\	
$	|	d_p^{[XY]}	|	$	&								&	$	8.9	\times	10^{-19}	$		&	BASE at Mainz ($p$ $g$)	\\	
$	|	d_p^{YY}	|	$	&								&	$	1.6	\times	10^{-18}	$		&	BASE at Mainz ($p$ $g$)	\\	
$	|	d_p^{ZX}	|	$	&								&	$	2.5	\times	10^{-18}	$		&	BASE at CERN ($\ol{p}$ $g$)	\\	
$	|	d_p^{ZY}	|	$	&								&	$	2.5	\times	10^{-18}	$		&	BASE at CERN ($\ol{p}$ $g$)	\\	
$	|	H_p^{TX}	|	$	&								&	$	2.3	\times	10^{-18}	$	GeV	&	BASE at CERN ($\ol{p}$ $g$)	\\	
$	|	H_p^{TY}	|	$	&								&	$	2.3	\times	10^{-18}	$	GeV	&	BASE at CERN ($\ol{p}$ $g$)	\\	
$	|	H_p^{TZ}	|	$	&								&	$	4.2	\times	10^{-19}	$	GeV	&	BASE at Mainz ($p$ $g$)	\\	
$	|	g_p^{TXT}	|	$	&								&	$	2.5	\times	10^{-18}	$		&	BASE at CERN ($\ol{p}$ $g$)	\\	
$	|	g_p^{TYT}	|	$	&								&	$	2.5	\times	10^{-18}	$		&	BASE at CERN ($\ol{p}$ $g$)	\\	
$	|	g_p^{TZT}	|	$	&								&	$	1.4	\times	10^{-18}	$		&	BASE at Mainz ($p$ $g$)	\\	
$	|	g_p^{XYX}	|	$	&								&	$	2.5	\times	10^{-18}	$		&	BASE at CERN ($\ol{p}$ $g$)	\\	
$	|	g_p^{XYY}	|	$	&								&	$	2.5	\times	10^{-18}	$		&	BASE at CERN ($\ol{p}$ $g$)	\\	
$	|	g_p^{XYZ}	|	$	&								&	$	3.6	\times	10^{-5}	$		&	BASE at CERN ($p-\ol{p}$ $c/m$)	\\	
$	|	g_p^{XZX}	|	$	&								&	$	1.8	\times	10^{-18}	$		&	BASE at CERN ($\ol{p}$ $g$)	\\	
$	|	g_p^{XZY}	|	$	&								&	$	8.1	\times	10^{-19}	$		&	BASE at Mainz ($p$ $g$)	\\	
$	|	g_p^{YZX}	|	$	&								&	$	8.1	\times	10^{-19}	$		&	BASE at Mainz ($p$ $g$)	\\	
$	|	g_p^{YZY}	|	$	&								&	$	1.8	\times	10^{-18}	$		&	BASE at CERN ($\ol{p}$ $g$)	\\					
\hline
\end{tabular}
\end{table*} 

\section{Prospects}
\label{prospects}

We provide in this section the prospects of improving the current SME limits 
or imposing first-time SME limits from time-variation analysis of
$\om_c^w$ and $\om_a^w$ of confined particles or antiparticles in
Penning-trap experiments. 
The limits listed in 
\Tab{limits} 
were obtained from pure sidereal-variation studies of the cyclotron and anomaly frequencies
and their comparisons between particles and antiparticles.
A detailed comparison of the SME coefficients 
from different rows in each of 
\Tab{cV}, \ref{cH}, \ref{aV}, and \ref{aH}
reveals that some SME coefficients 
that contribute to a pure sidereal variation also appear in a pure annual variation. 
This indicates that 
an additional pure annual variation study could potentially achieve greater sensitivities
to these overlapping SME coefficients than a pure sidereal variation analysis
due to the larger value of $\bE$ compared to $\beL$, 
by about two orders in magnitude.
Moreover,  
an additional pure annual variation or mix sidereal-annual variation 
analysis of $\om_c^w$ and $\om_a^w$ 
could constrain additional SME coefficients that 
are undetectable in a pure sidereal variation study.  

To give an explicit example, 
we consider the measurements of the anomaly frequencies
with a horizontal magnetic field,
corresponding to the setup 
used in Penning-trap experiments carried out by the BASE collaboration. 
As observed in 
\Sec{experiments}, 
the SME coefficients in 
$A_{0}^{(a,1)}$ and $\ol{A}_{0}^{(a,1)}$ 
can be constrained by particle-antiparticle comparison studies 
while those contributing to 
$A_{c}^{(a,1)}$, ${A}_{s}^{(a,1)}$, $A_{c2}^{(a,1)}$, and $A_{s2}^{(a,1)}$ 
are sensitive to pure sidereal variation studies.  
Some of the coefficients that contribute to these amplitudes 
in~\Tab{aH} also contribute to the $A_{C}^{(a,1)}$ and $A_{S}^{(a,1)}$ in the same table.  
These coefficients are identified as 
$b^T_w$, $H^{TX}_w$, $H^{TY}_w$, $d^{TT}_w$, $d^{ZX}_w$, $d^{ZY}_w$, 
$g^{TXT}_w$, $g^{TYT}_w$, $g^{XYX}_w$, and $g^{XYY}_w$.  
A pure annual variation study is potentially up to two orders of magnitude 
more sensitive to these coefficients because 
$\be_\oplus/\be_E\simeq 10^{2}$.   
In addition to the above SME coefficients, 
the pure annual variation analysis is also sensitive to two more SME coefficients, 
$d^{ZZ}_w$ and $g^{XYZ}_w$, 
that do not contribute to either the constant term 
or the amplitudes of the sidereal variations of the anomaly frequencies.  
We note that the coefficient $g^{XYZ}_w$ does contribute to the constant term $A_{0}^{(c,1)}$ of the cyclotron frequencies, 
and we used limit~\rf{cm-limit-base-cst}
to obtain a first bound on the size of this coefficient, 
which is listed in \Tab{limits}. 
However, 
a pure annual variation study of the anomaly frequencies could in principle
improve its limit from $10^{-5}$ level to $10^{-18}$ level, 
an improvement of 13 orders of magnitude, 
since the minimal SME contributions to the anomaly frequencies are independent of the term $|eB|$ 
compared to these to the cyclotron frequencies.  
There are four more SME coefficients, 
$d^{XZ}_w$, $d^{YZ}_w$, $g^{XZZ}_w$, and $g^{YZZ}_w$ 
that only contribute to the amplitudes of the mixed annual-sidereal variations 
in~\Tab{aH} 
and therefore, 
can only be detected by searching for this type of variation.  
These four coefficients are currently unconstrained in the proton sector. 
The attainable limits on their size based on a mixed annual-sidereal variation study 
are in the order of $10^{-18}$ or better.

\renewcommand{\arraystretch}{1.5}
\begin{table*}
\caption{
\label{access}
Accessibility of different time-variation studies to the SME coefficients.}
\setlength{\tabcolsep}{7pt}
\begin{tabular}{ccccc}
\hline
\hline		
Frequency	& Field &   Coefficients improvable                                                                         & New coefficients                                                             & New coefficients           \\
                   &       direction                  &   by an annual variation                                                                                                                                                &by an annual variation                                  &by a mixed variation   \\\hline
$\om_c^w$    & Vertical 	        & $c^{(TX)}_w$, $c^{(TY)}_w$, $g^{TXT}_w$, $g^{TYT}_w$, 	                                                & $b^T_w$, $c^{(TZ)}_w$, $g^{XYZ}_w$	                    & $g^{XZZ}_w$, $g^{YZZ}_w$\\
                   &                         & $g^{XYX}_w$, $g^{XYY}_w$, $g^{XZY}_w$, $g^{YZX}_w$                                                     &                                                                                       &\\
$\om_c^w$	& Horizontal       & $b^T_w$, $c^{(TX)}_w$, $c^{(TY)}_w$, $c^{(TZ)}_w$, $g^{TXT}_w$, $g^{TYT}_w$, 	         & None	                                                                     & $g^{XZZ}_w$, $g^{YZZ}_w$\\
                   &                         &  $g^{XYX}_w$, $g^{XYY}_w$, $g^{XZY}_w$, $g^{YZX}_w$, $g^{XYZ}_w$                             &                                                                                       &\\
$\om_a^w$     & Vertical 	        & $H^{TX}_w$, $H^{TY}_w$ , $d^{ZX}_w$,  $d^{ZY}_w$,                                                          & $b^T_w$, $d^{TT}_w$,$d^{ZZ}_w$, $g^{XYZ}_w$	 & $d^{XZ}_w$,$d^{YZ}_w$, $g^{XZZ}_w$, $g^{YZZ}_w$\\
                   &                         &  $g^{TXT}_w$,  $g^{TYT}_w$, $g^{XYX}_w$,  $g^{XYY}_w$	                                                &                                                                                      &\\
$\om_a^w$	& Horizontal       & $b^T_w$, $H^{TX}_w$, $H^{TY}_w$, $d^{TT}_w$, $d^{ZX}_w$,                                           &$d^{ZZ}_w$, $g^{XYZ}_w$	                                        & $d^{XZ}_w$,$d^{YZ}_w$, $g^{XZZ}_w$, $g^{YZZ}_w$\\
                   &                         &  $d^{ZY}_w$, $g^{TXT}_w$, $g^{TYT}_w$, $g^{XYX}_w$, $g^{XYY}_w$                               &                                                                                     &\\
\hline
\end{tabular}
\end{table*} 

To make the discussion complete, 
we provide 
\Tab{access}
 to summarize the improvable and new SME coefficients 
by additional annual and mixed sidereal-annual variation studies 
of the cyclotron and anomaly frequencies
for different field configurations. 
In this table,
the first column gives the frequencies of analysis 
and the second column displays the field direction used in an experiment. 
The third column lists the SME coefficients
that appear in both a pure sidereal variation (including the constant term) 
and a pure annual variation analysis. 
Note for these coefficients,
assuming a pure sidereal variation study has been performed already, 
an additional pure annual variation analysis has the advantage of improving their bounds by 
about two orders in magnitude,
as discussed at the beginning of this section.
The fourth column specifies the new SME coefficients that can be 
detected by an additional annual variation analysis.
The final column presents the additional SME coefficients 
that are sensitive to an additional mixed sidereal-annual variation study,
assuming pure sidereal and annual variation studies have been performed already. 
At the end of this section, 
we point out that since the limit of 
coefficient $g^{XYZ}_p$ 
listed in 
\Tab{limits}
is obtained by analyzing the cyclotron frequency 
difference between protons and antiprotons from 
the BASE experiment using a horizontal magnetic field, 
and this coefficient also lies in the category of 
``Coefficients improvable by an annual variation" for the 
case of $\om_c^w$ with a horizontal magnetic field in
\Tab{access},
an additional pure annual variation study could in principle improve its limit
by one or two orders of magnitude
due to the larger boost factor associated with the annual variation. 
The BASE collaboration recently performed an annual variation study 
of the cyclotron frequency difference 
between protons and antiprotons as a test of the Weak Equivalence Principle 
\cite{bo22}, 
so it has a great potential to improve the limit of $g^{XYZ}_p$.

\section{Summary}
\label{summary}

In this work,
we studied the Lorentz- and CPT-violating effects 
at linear boost order in Penning-trap experiments.
Within the SME framework,
we first reproduced the dominant Lorentz- and CPT-violating 
cyclotron and anomaly frequency shifts
of confined particles and antiparticles in Penning traps. 
We then presented a general discussion of transforming 
 SME coefficients from the apparatus frame to the Sun-centered frame
at linear boost order. 
Restricting the analysis to the minimal SME, 
the transformation was applied to express the cyclotron and anomaly frequency shifts 
in terms of the Sun-centered frame SME coefficients. 
We found that the expressions of these frequency shifts 
can be decomposed as a sum of harmonics of the Earth's sidereal frequency, 
the annual frequency, and the product of the two.
The amplitudes of the harmonics expressed 
in terms of the Sun-centered frame SME coefficients
were given in 
\Tab{cV},
\Tab{cH},
\Tab{aV},
and
\Tab{aH}.
Moving to the applications to Penning-trap experiments, 
we adopted the experimental measurements 
of the charge-to-mass ratios, the $g$ factors,
and their comparisons between protons and antiprotons 
from the ATRAP and BASE Penning-trap experiments
and translated them
in terms of the limits on the cyclotron and anomaly frequency shifts. 
Relating the frequency limits to the SME coefficients,
we extracted first-time constraints on 18 SME coefficients 
and improved the limits of 2 additional SME coefficients,
by about 4 and 10 orders of magnitude. 
The results were summarized in 
\Tab{limits}.
To conclude the work,
we provided some comments on improving the current SME limits
or imposing limits on new SME coefficients from 
different time-variation analysis. 
Following the present summary, 
Appendix~\ref{appA}
presented the Lorentz- and CPT-violating contributions 
to the cyclotron and anomaly frequency shifts
at the zeroth boost order.

Overall,
this work presents a general methodology for studying 
Lorentz- and CPT-violating boost effects in Penning-trap experiments. 
It provides a strong basis for future 
searches for Lorentz and CPT violation using Earth-based experiments.
Given the impressive measurement precision 
and excellent coverage of the SME coefficients,
Penning-trap experiments remain in the exciting category of experiments 
that have great potential to unveil novel signals 
for Lorentz and CPT violation in nature.

\appendix
\onecolumngrid
\section{Expressions for pure sidereal variations at both zeroth and first order in $\be$}
\label{appA}
The discussion in this work focused on Lorentz- and CPT-violating corrections 
to the cyclotron and anomaly frequencies at linear order in $\be$ 
due to the minimal SME terms.  
The results at the zeroth order in $\be$ including the nonminimal SME terms up to mass dimensions six
were given in 
Refs. \cite{dk16, dr20}. 
For completeness and the convenience of future time-variation studies of Penning-trap experiments,
we provide in this Appendix the full results including both the zeroth and linear order in $\be$.  
 
We start the discussion by reproducing the main results for the cyclotron frequency shifts obtained in 
Ref.~\cite{dr20}. 
The general form of the Lorentz- and CPT-violating shifts to the cyclotron frequencies of confined particles 
at the zeroth order in $\be$ is given by
\bea
\fr{\de \om_{c,0}^w}{eB}
&=&
A_0^{(c,0)}
+A_c^{(c,0)} \cos\om_\oplus T_\oplus+A_s^{(c,0)} \sin\om_\oplus T_\oplus
+A_{c2}^{(c,0)} \cos 2\om_\oplus T_\oplus+A_{s2}^{(c,0)} \sin 2\om_\oplus T_\oplus
\nn\\
&&
+A_{c3}^{(c,0)} \cos 3\om_\oplus T_\oplus+A_{s3}^{(c,0)} \sin 3\om_\oplus T_\oplus ,
\label{dwc0}
\eea
where $\om_\oplus$ is the sidereal frequency and  $T_\oplus$ the sidereal time.  
As discussed in 
\Sec{transformations},
the explicit expressions of the amplitudes $A_{\ast}^{(c,0)}$ in Eq.~\eqref{dwc0} 
depend on the field orientation in a trap. 
For a vertical magnetic field, 
$A_{\ast}^{(c,0)}$ are given by
 \bea
 A_0^{(c,0)} &=&\left(2 \bptw w Z c_\ch- m_w (\ctw w {XX}+\ctw w {YY})(1+c_\ch^2)-2 m_w( \ctw w {TT}+ \ctw w {ZZ} s_\ch^2 )\right)/2m_w^2-(\bftw w {ZXX}+\bftw w {ZYY} ) c_\ch c_{2\ch}\nn\\
 &&+(\bftw w {X(XZ)}+\bftw w {Y(YZ)}-\bftw w {ZZZ})c_\ch s^2_\ch, \nn\\
 A_c^{(c,0)} &=&\left(\bptw w X+2 m_w \ctw w {(XZ)}c_\ch - \fr{m_w^2}{8}\left(  \btw w {XYY} (7+ c_ {2\ch}) -  16\btw w {Z(XZ)} c^2_\ch+ \btw w {XXX} (5+ 3c_{2\ch}) \right)\right) \fr{s_\ch}{m_w^2}-( \btw w {XZZ}- \frac{1}{2} \btw w {Y(XY)})s^3_\ch, \nn\\
 A_s^{(c,0)} &=&\left(\bptw w Y+2 m_w \ctw w {(YZ)}c_\ch - \fr{m_w^2}{8}\left(  \btw w {YXX} (7+ c_ {2\ch}) -  16\btw w {Z(YZ)} c^2_\ch+ \btw w {YYY} (5+ 3c_{2\ch}) \right)\right) \fr{s_\ch}{m_w^2}-( \btw w {YZZ}- \frac{1}{2} \btw w {X(XY)})s^3_\ch, \nn\\
A_{c2}^{(c,0)} &=&  \left(\ctw w {XX} - \ctw w {YY}-m_w\left( 2 \btw w {Y(YZ)} - 2\btw w {X(XZ)}  - \btw w {ZXX} + \btw w {ZYY} \right) c_\ch \right)  \fr{s^2_\ch}{2 m_w},\nn\\
A_{s2}^{(c,0)} &=&  \left(\ctw w {(XY)}  + m_w\left( \btw w {Y(XZ)}+ \btw w {X(YZ)}  + \btw w {Z(XY)}  \right) c_\ch \right)  \fr{s^2_\ch}{m_w},\nn\\
A_{c3}^{(c,0)} &=&- \fr{s_\ch^3}{4}\left( 2 \btw w {Y(XY)}-\btw w {XXX}+ \btw w {XYY}\right),\hskip 5pt   A_{s3}^{(c,0)}  =
 \fr{s_\ch^3}{4}\left( 2 \btw w {X(XY)}+\btw w {YXX}- \btw w {YYY}\right).
 \label{DcV}
 \eea
For a horizontal magnetic field, 
we have
 \bea
 A_{0}^{(c,0)} 
 &=&
 -\left[
 \bptw w {Z} c_\th s_\ch+m_w \left( \ctw w {TT}+  \half(\ctw w {XX} + \ctw w {YY}) (c^2_\th + c^2_\ch s^2_\th + s^2_\ch ) + \ctw w {ZZ} (c^2_\ch + s^2_\th s^2_\ch) \right)
  \right]/m_w^2
 \nn\\
 &&
 -\frac{1}{8} (\btw w {X(XZ)} + \btw w {Y(YZ)} - \btw w {ZZZ}) c_\th(5 s_ \chi -4 c_ {2\th} s^3_ \chi + s_{3\chi})- \frac{1}{16} (\btw w {ZXX} + \btw w {ZYY}) \left(2 c^3_\th s_{3\chi} - c_\th s_\ch (3 c_{2\th} + 11 )\right),
 \nn\\
 A_c^{(c,0)} 
 &=&
 \left( \bptw w {X} c_\th c_\ch + \bptw w {Y} s_\th -m_w(\ctw w {(XZ)} c^2_\th s_{2\ch} + \ctw w {(YZ)} s_{2\th} s_\ch)\right)/m_w^2- \frac{1}{16} \btw w {Y(XY)} c_\th\left((c_{2\th}-7) c_\chi-2 c^2_\th c_{ 3 \chi}\right) 
 \nn\\
  &&
  -c_\th c_\chi \frac{1}{16}\left(\btw w {XXX} ( 3 c_{2 \th}-6 c^2_{\th} c_{2 \chi} +7)+\btw w {XYY} ( c_{2\th}-2 c_{\th}^2c_{2\chi}+13)+16\,\btw w {XZZ} (s^2_\th s^2_\chi+c^2_\chi) -16\,\btw w {Z(XZ)} c^2_\th s_{2\chi} \right
  )\nn\\
  &&
  + \frac{1}{2}s_\th\left( \btw w {X(XY)} (c^2_\th c_{2\chi} +s^2_\th c^2_\chi +s^2_\chi)-2\btw w {YZZ} (s^2_\th s^2_\chi+c^2_\chi)+4\btw w {Z(YZ)} c^2_\th s^2_\chi \right)
  \nn\\
  &&
  -\frac{1}{32}\left[ \btw w {YXX} \left(s_\theta(25-4 c^2_\th c_{2\chi} )+s_{3\th}\right)+\btw w {YYY} \left(s_\th(11-12 c^2_\th c_{2\chi})+3 s_{3\th}\right)\right],\nn\\
  A_s^{(c,0)} 
  &=&
  \left(  \bptw w {Y} c_\th c_\ch- \bptw w {X} s_\th +m_w( \ctw w {(XZ)} s_{2\th} s_\ch - \ctw w {(YZ)}  c^2_\th s_{2\ch}  )\right)/m_w^2+2c^2_\th s^2_\chi(\btw w {Z(YZ)} c_\th c_\chi -\btw w {Z(XZ)} s_\th)
  \nn\\
 && 
 -\frac{1}{16} c_\th c_\chi \left[ \btw w {YXX} (c_{2\th}-2 c^2_\th c_{2\chi} +13)+\btw w {YYY} (3 c_{2\th}-6 c^2_{\th}c_{2\chi} +7)+16\btw w {YZZ} (s^2_\th s^2_\chi+c^2_\chi)\right]
 \nn\\
&& 
-\frac{1}{32}\left[ 2\btw w {X(XY)} c_\th\left((c_{2\th}-7) c_\chi-2 c^2_\th c_{3\chi}\right)+ \btw w {XXX} \left(s_\th(12 c^2_\th c_{2\chi} -11)-3 s_{3\th}\right)+\btw w {XYY} \left(s_\th(4 c^2_\th c_{2\chi} -25)-s_{3 \th}\right)\right]
\nn\\
&&
+s_\th\left[\btw w {XZZ} (s^2_\th s^2_\chi+c^2_\chi)- \frac{1}{2} \btw w {Y(XY)}(c^2_\th c_{2\chi} +s^2_\th c^2_\chi+s^2_\chi)\right],
\label{DcH1}
\eea
and
\bea
A_{c2}^{(c,0)} 
&=&
-\left(\frac{1}{8} (\ctw w {XX} -  \ctw w {YY}) (1-3c_{2\th} - 2c^2_\th c_{2\ch}) - \ctw w {(XY)} c_\ch s_{2\th}\right)/m_w-(\btw w {X(YZ)} +\btw w {Y(XZ)} +\btw w {Z(XY)} )s_\th c^2_\th s_{2\chi}
\nn\\
&&
+\frac{1}{32} \left(2 \btw w {X(XZ)} -2 \btw w {Y(YZ)}+\btw w {ZXX} -\btw w {ZYY}\right) \left((c_\th-5c_{3\th}) s_\chi -4 c^3_\th  s_{3 \chi} \right),
\nn\\
A_{s2}^{(c,0)} 
&=&
-\left(  \half (\ctw w {XX} -  \ctw w {YY}) c_\ch s_{2\th} + \frac{1}{4} \ctw w {(XY)} (1-3c_{2\th} - 2c^2_\th c_{2\ch})\right)/m_w+\left(\btw w {X(XZ)} -\btw w {Y(YZ)} + \half (\btw w {ZXX} - \btw w {ZYY})\right) s_\th c^2_\th s_{2\chi}
\nn\\
&&
+ \frac{1}{16} \left(\btw w {X(YZ)} + \btw w {Y(XZ)} + \btw w {Z(XY)} \right) \left((c_\th -5 c_{3 \th}) s_\chi -4 c^3_{\th} s_ {3 \chi}\right),
\nn\\
A_{c3}^{(c,0)} 
&=&
\frac{1}{64}\left( \btw w {XYY}-\btw w {XXX} +2 \btw w {Y(XY)}\right)\left(3 (c_\th -5 c_{3 \th} ) c_\chi -4 c^3_\th  c_{3 \chi} \right)
\nn\\
&&
- \frac{1}{32}\left(2\btw w {X(XY)} +  \btw w {YXX} -\btw w {YYY}\right) \left(3s_\th (1-4 c^2_\th  c_{2 \chi}) - 5 s_{3 \th} \right),
 \nn\\
A_{s3}^{(c,0)} 
&=& 
\frac{1}{32} \left(\btw w {XXX}-\btw w {XYY}  -2 \btw w {Y(XY)}\right) \left(3s_\th (1-4 c^2_\th  c_{2 \chi}) - 5 s_{3 \th} \right)
\nn\\
&&
 -\frac{1}{64} \left( 2\btw w {X(XY)} +  \btw w {YXX} -\btw w {YYY}\right) \left(3 (c_\th -5 c_{3 \th} ) c_\chi -4 c^3_\th  c_{3 \chi} \right),
\label{DcH2}
\eea
where the definitions of all the relevant tilde effective coefficients can be found in 
Ref.~\cite{dr20}. 
The frequency shifts $\de \om_{c,0}^{\ol w} $ for antiparticles can be obtained by replacing the amplitudes 
$A_{\ast}^{(c,0)}$ by  ${\ol A}_{\ast}^{(c,0)}$ in Eq.~\eqref{dwc0}.  
The expressions of amplitudes ${\ol A}_{\ast}^{(c,0)}$ are obtained from Eqs. \eqref{DcV}, \eqref{DcH1}, and \eqref{DcH2}
by replacing $\bptw w J \rightarrow-\bptws w J$, $\ctw w {JK}\rightarrow\ctws w {JK}$, and  $\bftw w {JKL} \rightarrow-\bftws w {JKL}$.

For pure sidereal variation studies of the cyclotron frequencies,
the expressions of the different amplitudes can be obtained by summing 
Eqs. \rf{dwc0} and \rf{dwc1}, 
given by
\bea
| \de \om_c^{w}|_{\rm{const}}
&=&
|A^{(c,0)}_{0}+A^{(c,1)}_{0}|,
\nn\\
| \de \om_c^{w}|_{\rm{1st}}
&=&
\sqrt{ \left(A^{(c,0)}_c+A^{(c,1)}_c\right)^2+\left(A^{(c,0)}_s+A^{(c,1)}_s\right)^2 }, 
\nn\\
| \de \om_c^{w}|_{\rm{2nd}}
&=&
\sqrt{ \left(A^{(c,0)}_{c2}+A^{(c,1)}_{c2}\right)^2+\left(A^{(c,0)}_{s2}+A^{(c,1)}_{s2}\right)^2 },
\nn\\
 | \de \om_c^{w}|_{\rm{3rd}}
 &=&
 \sqrt{\left(A^{(c,0)}_{c3}\right)^2+\left(A^{(c,1)}_{s3}\right)^2 },
\label{wcSidApp}
\eea
where the $A^{(c,1)}_{\ast}$ amplitudes are given by Tables \ref{cV} and \ref{cH}.  
Note the constraining of amplitudes 
$|A^{(c,0)}_{0}|$ and $|A^{(c,1)}_{0}|$ in $| \de \om_c^{w}|_{\rm{const}}$  
requires comparisons of different cyclotron frequencies.  

For cyclotron frequency shifts comparisons $| \de \om_c^{w}- \de \om_c^{\ol w}|$
between particles and antiparticles, 
the expressions of different amplitudes for a pure sidereal variation study are given by
 \bea
| \de \om_c^{w}- \de \om_c^{\ol w}|_{\rm{const}}
&=&
|A^{(c,0)}_{0}+A^{(c,1)}_{0}-{\ol A}^{(c,0)}_{0}-{\ol A}^{(c,1)}_{0}|,
\nn\\
| \de \om_c^{w}- \de \om_c^{\ol w}|_{\rm{1st}}
&=&
\sqrt{ \left(A^{(c,0)}_c+A^{(c,1)}_c-{\ol A}^{(c,0)}_c-{\ol A}^{(c,1)}_c\right)^2
+\left(A^{(c,0)}_s+A^{(c,1)}_s-{\ol A}^{(c,0)}_s-{\ol A}^{(c,1)}_s\right)^2 }, 
\nn\\
|\de \om_c^{w}- \de \om_c^{\ol w}|_{\rm{2nd}}
&=&
\sqrt{ \left(A^{(c,0)}_{c2}+A^{(c,1)}_{c2}-{\ol A}^{(c,0)}_{c2}-{\ol A}^{(c,1)}_{c2}\right)^2
+\left(A^{(c,0)}_{s2}+A^{(c,1)}_{s2}-{\ol A}^{(c,0)}_{s2}-{\ol A}^{(c,1)}_{s2}\right)^2 },
\nn\\
 | \de \om_c^{w}- \de \om_c^{\ol w}|_{\rm{3rd}}
 &=&
 \sqrt{\left(A^{(c,0)}_{c3}-{\ol A}^{(c,0)}_{c3}\right)^2
 +\left(A^{(c,1)}_{s3}-{\ol A}^{(c,1)}_{s3}\right)^2 } .
\label{wcSidAppCp}
\eea
The limits 
\rf{cm-limit-atrap},
\rf{cm-limit-base-cst},
and
\rf{cm-limit-base-1st}
obtained by pure sidereal variation studies of $\om_c$ 
can be used to impose constraints on the first two expressions in
\eq{wcSidAppCp}.
This expands the results obtained in 
Ref.~\cite{dr20} 
by including the contributions at linear order in $\be$ contained 
in the amplitudes $A^{(c,1)}_{\ast}$ and ${\ol A}^{(c,1)}_{\ast}$.

Moving the discussion to the anomaly frequencies ,
the general form of the Lorentz- and CPT violating shifts to the anomaly frequencies
at the leading order in $\be$ is given by
\beq
\de \om_{a,0}^w
=
A_0^{(a,0)}
+A_c^{(a,0)} \cos\om_\oplus T_\oplus+A_s^{(a,0)} \sin\om_\oplus T_\oplus
+A_{c2}^{(a,0)} \cos 2\om_\oplus T_\oplus+A_{s2}^{(a,0)} \sin 2\om_\oplus T_\oplus,
\label{dwa0}
\eeq
with the amplitudes $A_{\ast}^{(a,0)}$ given by
\bea
A_0^{(a,0)} 
&=& 
2\btw w Z c_\ch -B\left(2\bftw w {ZZ}+(\bftw w {XX} +\bftw w {YY} -2\bftw w {ZZ}) s_\ch^2\right), 
\nn\\ 
A_{c}^{(a,0)} 
&=& 
2\btw w X s_{\ch} - 2 B\, \bftw w {(XZ)} s_{2\chi},
\nn\\ 
A_{s}^{(a,0)} 
&=& 
2\btw w Y s_\ch - 2 B\, \bftw w {(YZ)} s_{2\chi}, 
\nn\\ 
A_{c2}^{(a,0)} 
&=& 
B(\bftw w {YY} - \bftw w {XX})s_\ch^2, 
\nn\\ 
A_{s2}^{(a,0)} 
&=& 
-2B\,\bftw w {(XY)} s_\ch^2,
\label{DaV}
\eea
for a vertical magnetic field,
and  
\bea
A_0^{(a,0)} 
&=& 
-2\btw w Z s_\ch c_\th -B\left((\bftw w {XX} +\bftw w {YY})(c_\ch^2 c_\th^2+s_\th^2)+2 \bftw w {ZZ} c_\th^2 s_\ch^2 \right), 
\nn \\
A_{c}^{(a,0)} 
&=& 
2(\btw w X c_\th c_\ch +\btw w Y s_\th)+ 4 B\,c_\th s_\ch\left(\bftw w {(XZ)} c_\th c_\ch +\bftw w {(YZ)} s_\th \right),
 \nn\\
A_{s}^{(a,0)} 
&=& 
2(\btw w Y c_\th c_\ch -\btw w X s_\th)+ 4 B\,c_\th s_\ch\left(\bftw w {(YZ)} c_\th c_\ch -\bftw w {(XZ)} s_\th \right), 
\nn \\
A_{c2}^{(a,0)} 
&=& 
 B \left((\bftw w {YY}-\bftw w {XX})(c^2_\th c^2_\ch-s_\th^2) -2 \bftw w {(XY)} c_\ch s_{2\th} \right),
 \nn \\
A_{s2}^{(a,0)} 
&=& 
 -B \left(2 \bftw w {(XY)}(c^2_\th c^2_\ch-s_\th^2) +(\bftw w {YY}-\bftw w {XX}) c_\ch s_{2\th} \right),
\label{DaH}
\eea
for a horizontal magnetic field.
Similarly, 
the anomaly frequency shifts $\de \om_{a,0}^{\ol w}$ for antiparticles 
can be determined by replacing $A_{\ast}^{(a,0)}$ to ${\ol A}_{\ast}^{(a,0)}$
in Eq.~\eqref{dwa0},
with ${\ol A}_{\ast}^{(a,0)}$ given by 
replacing $\btw w J \rightarrow-\btws w J$ and $\bftw w {JK} \rightarrow-\bftws w {JK}$
in Eqs.~\eqref{DaV} and \eqref{DaH}.

The corresponding different amplitudes in a 
pure sidereal variation study of the anomaly frequencies
can be determined by adding 
\eq{dwa1} to \rf{dwa0}.
These amplitudes were found to be
\bea
| \de \om_a^{w}|_{\rm{const}}
&=&
|A^{(a,0)}_{0}+A^{(a,1)}_{0}|,
\nn\\
| \de \om_a^{w}|_{\rm{1st}}
&=& 
\sqrt{ \left(A^{(a,0)}_c+A^{(a,1)}_c\right)^2+\left(A^{(a,0)}_s+A^{(a,1)}_s\right)^2 }, 
 \nn \\
| \de \om_a^{w}|_{\rm{2nd}}
&=& 
\sqrt{ \left(A^{(a,0)}_{c2}+A^{(a,1)}_{c2}\right)^2+\left(A^{(a,0)}_{s2}+A^{(a,1)}_{s2}\right)^2 },
\label{waSidApp}
\eea
where the amplitudes $A^{(a,1)}_{\ast}$ are given by Table~\ref{aV} for a vertical magnetic field 
and by Table~\ref{aH} for a horizontal one. 
Applying the limits 
\rf{g-limit-base-1st} and \rf{g-limit-base-2nd}
from pure sidereal variation studies of $\om_a$
to the expressions of 
$| \de \om_a^{w}|_{\rm{1st}}$ and $| \de \om_a^{w}|_{\rm{2nd}}$
in Eq.~\eqref{waSidApp} expands the results in 
Ref.~\cite{dk16} 
by including amplitudes $A^{(a,1)}_{\ast}$ from linear order boost contributions.

For particle-antiparticle anomaly frequency shifts comparisons 
$| \de \om_a^{w}- \de \om_a^{\ol w}|$,
the corresponding amplitudes are found in a similar way,
 \bea
| \de \om_a^{w}- \de \om_a^{\ol w}|_{\rm{const}}
&=&
|A^{(a,0)}_{0}+A^{(a,1)}_{0}-{\ol A}^{(a,0)}_{0}-{\ol A}^{(a,1)}_{0}|,
\nn\\
| \de \om_a^{w}- \de \om_a^{\ol w}|_{\rm{1st}}
&=& 
\sqrt{ \left(A^{(a,0)}_c+A^{(a,1)}_c-{\ol A}^{(a,0)}_c-{\ol A}^{(a,1)}_c\right)^2
+\left(A^{(a,0)}_s+A^{(a,1)}_s-{\ol A}^{(a,0)}_s-{\ol A}^{(a,1)}_s\right)^2 }, 
 \nn \\
| \de \om_a^{w}- \de \om_a^{\ol w}|_{\rm{2nd}}
&=& 
\sqrt{ \left(A^{(a,0)}_{c2}+A^{(a,1)}_{c2}-{\ol A}^{(a,0)}_{c2}-{\ol A}^{(a,1)}_{c2}\right)^2
+\left(A^{(a,0)}_{s2}+A^{(a,1)}_{s2}-{\ol A}^{(a,0)}_{s2}-{\ol A}^{(a,1)}_{s2}\right)^2 }.
\label{waSidAppCp}
\eea
We note that if the anomaly frequency comparison $| \de \om_a^{w}- \de \om_a^{\ol w}|$
is based on \eq{gratio-lv} using experiments with magnetic fields of different strengths,
we can define a factor $\xi \equiv \om_c^{w}/\om_c^{\ol w}$ to represent the cyclotron frequency ratio  
between particles and antiparticles. 
To obtain limits of the SME coefficients,
this factor is then incorporated into the comparison as 
$| \de \om_a^{w}- \xi \de \om_a^{\ol w}|$
and carried along with the amplitudes 
${\ol A}^{(a,0)}_{\ast}$ and ${\ol A}^{(a,1)}_{\ast}$
in \eq{waSidAppCp} as well. 
Using the limit \rf{g-limit-basecom-cst} together with the first expression in
\eq{waSidAppCp},
we can extend the results obtained in 
Ref.~\cite{dk16}
by including the contributions at linear order in $\be$.

\twocolumngrid

\end{document}